\documentclass[prd, twocolumn, preprintnumbers, superscriptaddress, nofootinbib]{revtex4}%
\pdfoutput=1 

\usepackage{amssymb,hyperref, amsmath}
\usepackage{color}
\usepackage{graphicx}
\usepackage{multirow}

\begin{document}

\bibliographystyle{apsrev}

\preprint{JLAB-THY-11-1387}

\title{The lightest hybrid meson supermultiplet in QCD}

\author{Jozef J. Dudek}
\email{dudek@jlab.org}
\affiliation{Jefferson Laboratory, 12000 Jefferson Avenue,  Newport News, VA 23606, USA}
\affiliation{Department of Physics, Old Dominion University, Norfolk, VA 23529, USA}

\begin{abstract}
We interpret the spectrum of meson states recently obtained in non-perturbative lattice QCD calculations in terms of constituent quark-antiquark bound states and states, called `hybrids', in which the $q\bar{q}$ pair is supplemented by an excitation of the gluonic field. We identify a lightest supermultiplet of hybrid mesons with $J^{PC} = (0,1,2)^{-+}, 1^{--}$ built from a gluonic excitation of chromomagnetic character coupled to $q\bar{q}$ in an $S$-wave. The next lightest hybrids are suggested to be quark orbital excitations with the same gluonic excitation, while the next distinct gluonic excitation is significantly heavier. Existing models of gluonic excitations are compared to these findings and possible phenomenological consequences explored.
\end{abstract}


\maketitle 

\section{Introduction}
Models of hadrons as bound states of minimal numbers of ``constituent" or ``dressed" quarks \cite{Godfrey:1985xj,Chang:2011ei} can be motivated by the description of the flavor and $J^{P(C)}$ systematics of experimental meson and baryon spectra \cite{Nakamura:2010zzi}. In particular, the absence of meson states with isospin greater than 1 or magnitude of strangeness greater than 1 suggests a picture in which mesons are constituent quark-antiquark states where these constituent quarks have the same flavor quantum numbers as the quarks in the QCD Lagrangian. Higher Fock-state configurations, such as $qq\bar{q}\bar{q}$, would in general give rise to flavor configurations not observed experimentally. Furthermore, all well-established states have $J^{PC}$ within the set $0^{-+}, 0^{++}, 1^{--}, 1^{+-}, 1^{++}, 2^{-+}, 2^{++} \ldots$, all of which are accessible to a fermion-antifermion pair with orbital angular momentum, $L$. This suggests that the constituent quark degrees-of-freedom have spin-$\tfrac{1}{2}$ and the ordering of states in the spectrum implies increasing energy with increasing $L$. There is some theoretical support for the idea that constituent quarks with effective mass (at low energy scales) of several hundred MeV arise out of the almost massless quarks of the QCD Lagrangian in the process of spontaneous chiral symmetry breaking (see e.g. \cite{Vogl:1991qt, Szczepaniak:2000bi, Roberts:2007ji}).

Although the gross features of the experimental spectrum strongly suggests this $q\bar{q}$ structure assignment for most states, such simplicity looks peculiar viewed from the perspective of QCD, in which quarks couple strongly to a self-interacting gluonic field. From the generic properties of QCD we might expect to have states in which the gluonic field itself is excited and carries $J^{PC}$ quantum numbers. Absent any `valence' quark content, such a state is called a `glueball' and these basis states would be expected to appear mixed into the spectrum of isoscalar mesons \cite{Crede:2008vw}. The addition of a constituent quark-antiquark pair to an excited gluonic field gives us what we term a `hybrid' meson, where these states have the flavor quantum numbers accessible to $q\bar{q}$. They do not simply augment the regular $q\bar{q}$ spectrum however since the excited gluonic field could carry $J_g^{P_gC_g}$ quantum numbers other than $0^{++}$. The gluonic quantum numbers can couple to the $q\bar{q}$ quantum numbers to give rise to so-called `exotic' meson $J^{PC}$, literally those not accessible to a $q\bar{q}$ pair alone. Observation of a state with quantum numbers in this set, $0^{--}, 0^{+-}, 1^{-+}, 2^{+-}, 3^{-+}\ldots$, is considered a smoking gun signature for states beyond the simple $q\bar{q}$ assignment.

Now since the constituent $q\bar{q}$ picture was largely motivated by the \emph{absence} of such exotic $J^{PC}$ states in the observed spectrum, we must question why they have not been observed. Possibilities include a large energy scale associated with gluonic excitations, that places the states in a high mass region that has not been explored. Alternatively, the mass scale may be modest, but the states may have production/decay characteristics sufficiently different from $q\bar{q}$ mesons that experiments have thus far missed them \cite{Isgur:1985vy, Meyer:2010ku}. 

At present the experimental situation surrounding exotic $J^{PC}$ mesons is confused, with $1^{-+}$ states being been claimed and disclaimed in several final states (see the review in \cite{Meyer:2010ku}), while other exotic $J^{PC}$ are essentially unexplored. There is hope that in the current decade we will see new data from GlueX, CLAS12, BESIII, Compass and PANDA that, through properly constrained analysis, will indicate definitively the presence or absence of exotic $J^{PC}$ states in the light-meson spectrum.

From the theoretical side, estimates of hybrid meson properties have traditionally followed from models which proceed from an assumed form for the gluonic excitation. There are a spread of such models and they make divergent predictions for the spectrum that, in the absence of clear experimental results, cannot be tested. Lattice QCD offers the possibility of computing meson properties directly from QCD, through numerical computation of the QCD path integral under the controlled approximation of a finite, discretised, space-time grid. While lattice calculations have made impressive progress in precision computation of the masses of the lightest hadrons\cite{Durr:2008zz}, it is only very recently that extraction of excited state properties has become feasible \cite{Dudek:2007wv,Dudek:2009qf,Dudek:2010wm,Dudek:2011tt, Edwards:2011jj}.

In this paper we show that the lattice QCD spectroscopy data presented in \cite{Dudek:2009qf,Dudek:2010wm,Dudek:2011tt} can be interpreted in terms of a spectrum of constituent $q\bar{q}$ states supplemented by hybrid mesons, with both exotic and non-exotic $J^{PC}$. The lightest supermultiplet of hybrid mesons is clearly identified for the first time in a framework directly connected to non-perturbative QCD. We compare the systematics of the extracted spectrum with expectations of existing models of gluonic excitations and begin to develop a QCD-motivated phenomenology that may be compared to experimental data.


\section{Gluonic Excitation Models}
\label{models}

A number of models proposing possible forms that the gluonic excitation within a hybrid meson might take have been explored \cite{Horn:1977rq, Isgur:1984bm, Barnes:1982tx, Chanowitz:1982qj, General:2006ed, Guo:2008yz}. Notably, they make differing predictions for the spectrum of hybrid mesons, so there is a hope of eventually distinguishing which, if any of them, resembles reality.

In the flux-tube model \cite{Isgur:1984bm}, the gluonic field in a meson is assumed to form itself into a tube between the quark and antiquark. Motivation comes from the strong coupling limit of QCD \cite{Kogut:1974ag}, explicit computation of the action density between well-separated static color sources \cite{Bali:1994de, Bissey:2009gw} and the corresponding static potentials \cite{Juge:2002br, Bali:2003jq}. The model confines quarks and gives rise to linear Regge trajectories for conventional mesons. Hybrid mesons are described in terms of transverse oscillations of the tube, and the high degree of symmetry possessed by the system leads to a large degeneracy for the predicted lightest hybrid supermultiplet\footnote{exotic $J^{PC}$ shown in bold face}: $(0,\mathbf{1},2)^{-+},(\mathbf{0},1,\mathbf{2})^{+-}, 1^{--}, 1^{++}$. The model can be extended by including flux-tube \emph{breaking} dynamics that give rise to meson decay \cite{Kokoski:1985is}.

Bag model approaches \cite{Barnes:1982tx,Chanowitz:1982qj} confine quarks inside a cavity. Gluonic excitations correspond to gluonic field modes allowed by the boundary conditions on the cavity wall and it is found that a ``TE" mode with $J^{P_gC_g}_g=1^{+-}$ is lightest, significantly lighter than a ``TM" mode with $J^{P_gC_g}_g=1^{--}$. This gives rise to a lightest hybrid supermultiplet $(0,\mathbf{1},2)^{-+},1^{--}$ if the $q\bar{q}$ is in an $S$-wave. Perturbative elaborations of the model allow for non-exotic $J^{PC}$ hybrid states to mix with $q\bar{q}$ and for states to decay to pairs of lighter mesons. 

Another alternative description of gluonic excitations proposes that they correspond to the addition to the $q\bar{q}$ pair of one or more massive constituent gluons. By analogy to constituent quarks being massive excitations of a non-trivial vacuum state filled with $q\bar{q}$ pairs, constituent gluons would be quasi-particle excitations of a non-trivial gluonic vacuum state. Particular implementations of this idea include the Coulomb-gauge many-body formalism in which the transverse quasi-gluons are excitations of a variationally obtained approximate non-perturbative vacuum state \cite{Szczepaniak:2000bi, General:2006ed, Guo:2008yz}.

In the simplest constituent gluon picture one can add a $J_g^{P_gC_g}=1^{--}$ transverse quasi-gluon in a relative $S$-wave with respect to a $q\bar{q}$ pair. This gives rise to a supermultiplet $(0,1,2)^{++},1^{+-}$ if the $q\bar{q}$ is in an $S$-wave. This  exactly resembles a $q\bar{q}$ $P$-wave and would be very hard to pick out of a spectrum unless some distinguishing decay characteristics could be observed. With the $q\bar{q}$ pair in a $P$-wave however, one has a large supermultiplet $\mathbf{0^{--}}, (1^{--})^3, (2^{--})^2, 3^{--}, 0^{-+}, \mathbf{1^{-+}}, 2^{-+}$, featuring negative parity exotic states. Notably there is a light $0^{--}$ exotic - a prediction essentially unique to this model \cite{General:2006ed, Horn:1977rq}.

There are arguments based upon the existence of repulsive three-body ($q\bar{q}g$) forces \cite{Guo:2008yz} and the behavior of adiabatic potentials in QCD at short distances \cite{Bali:2003jq}, that in fact having the quasi-gluon in a $P$-wave with respect to the $q\bar{q}$, coupled to total gluon spin $J_g^{P_gC_g} = 1^{+-}$, is energetically favorable. In this case one has a lightest hybrid supermultiplet of $(0,\mathbf{1},2)^{-+}, 1^{--}$ with the $q\bar{q}$ pair in an $S$-wave and, somewhat heavier,  a larger supermultiplet of $\mathbf{0^{+-}}, (1^{+-})^3, (\mathbf{2^{+-}})^2, 3^{+-}, (0,1,2)^{++}$ where the $q\bar{q}$ are in a $P$-wave.

Although they make very different predictions for the spectrum, without experimental evidence for a pattern of hybrid mesons it is difficult to favor any one of these models above the others. It is here that we will take advantage of recent improvements in spectroscopic calculations in lattice QCD to provide solid theoretical evidence for a spectrum of hybrid mesons with both exotic and non-exotic $J^{PC}$.


\section{Lattice QCD Spectroscopy}

Lattice QCD offers a means to obtain hadronic quantities directly from QCD by computing  appropriate gauge-invariant correlation functions constructed from quark and gluon fields. By considering it on a finite grid, the QCD path integral can be Monte-Carlo sampled and correlation functions determined with finite statistical precision. In principal one can, by performing multiple calculations, extrapolate in decreasing lattice spacing, and incresing finite volume of the box - in this sense the lattice approximation is systematically improvable. There are technical challenges, notably the poor scaling of the computational cost with decreasing quark mass, which means that most calculations currently have to be performed with unphysically heavy up and down quark masses. 

The spectrum of hadrons follows most directly from two-point correlation functions,
\begin{equation}
C_{ij}(t) = \big\langle	0 \big|   {\cal O}_i(t) {\cal O}_j(0)   \big| 0 \big\rangle \nonumber
\end{equation}
where ${\cal O}$ are gauge-invariant combinations of quark and gluon fields (expressed via the parallel transporters, or ``links" of the lattice) that have the desired hadron quantum numbers. Detailed discussion of the methodology required to extract an excited state spectrum with determined continuum $J^{PC}$ quantum numbers can be found in \cite{Dudek:2010wm} - here we summarise just the salient points. The quantities extracted from the calculations are the state masses, $m_\mathfrak{n}$, and vacuum-operator-state ``overlaps", $Z^\mathfrak{n}_i \equiv \big\langle \mathfrak{n} \big| {\cal O}_i \big| 0\big\rangle$, that appear in a spectral decomposition of a two-point correlation function, 
\begin{equation}
 C_{ij}(t) = \big\langle 0 \big| {\cal O}_i(t) {\cal O}_j(0) \big| 0\big\rangle = \sum_\mathfrak{n} \tfrac{Z_i^\mathfrak{n} Z_j^\mathfrak{n}}{2 m_\mathfrak{n}} \, e^{-m_\mathfrak{n} t}, \nonumber
\end{equation}
where the sum is over all eigenstates, $|\mathfrak{n}\rangle$, of the QCD Hamiltonian in finite volume with the same quantum numbers as the operators ${\cal O}_{i,j}$. There is a variational method of analysis \cite{Michael:1985ne,Luscher:1990ck} that analyses a \emph{matrix} of such correlators built using a basis of operators, returning best estimates for masses and overlaps.

In \cite{Dudek:2010wm}, a large operator basis for mesons was built using fermion bilinears projected onto zero meson momentum, featuring up to three gauge-covariant derivatives\footnote{The ``forward-backward" gauge-covariant derivative $\overleftrightarrow{D}_\mu = \overleftarrow{\partial}_\mu - \overrightarrow{\partial}_\mu - 2 i g A_\mu$ makes construction of operators of definite charge-conjugation relatively simple. On a cubic lattice the derivatives are implemented via parallel-transported finite-differences featuring the $SU(3)$ matrices living on the links. } ($d=0,1,2,3$),
\begin{equation}
\sum_{\vec{x}} \big\langle J_\Gamma m_\Gamma; J_D m_D \big| J, M \big\rangle \Big[ \bar{\psi} \Gamma_{m_\Gamma} D^{[d]}_{J_D,m_D} \psi\Big](\vec{x}),\nonumber
\end{equation}
where
\begin{align}
D^{[1]}_{J_D=1,m} &= \vec{\epsilon}(m) \cdot \overleftrightarrow{D}, \nonumber \\
D^{[2]}_{J_D=\{0,1,2\},m} &= \big\langle 1, m_1; 1, m_2 \big|J_D, m \big\rangle D^{[1]}_{J_D=1,m_1} D^{[1]}_{J_D=1,m_2}, \nonumber 
\end{align}
and where the three-derivative construction can be found in \cite{Dudek:2010wm}. The use of a circular polarisation basis, $\vec{\epsilon}\,(m)$, for the derivatives enables the use of simple $SO(3)$ Clebsch-Gordan coefficients for the angular momentum constructions.

The quark fields featuring in these expressions do not just exist on a single site on the lattice - they are smeared over space in a gauge-invariant way. The particular form of smearing used is ``distillation"\cite{Peardon:2009gh}, which proves to be a highly efficient way to compute a large number of correlators with operators that sample dominantly the lowest energy modes of QCD relevant to low-lying hadron states. Furthermore all gauge-fields entering into the operator construction are smeared in a gauge-covariant manner known as ``stout-smearing"\cite{Morningstar:2003gk} that preserves transformational properties of the original fields. The consistent use of gauge-invariant constructions is relevant since the computation is not performed in any fixed gauge.

The calculations reported in \cite{Dudek:2010wm} were performed on dynamical anisotropic Clover lattices with three flavors of quark, the lightest two of which are mass degenerate and the third is tuned to describe the strange quark. The lattice spacing in the spatial directions is $a_s \sim 0.12 \,\mathrm{fm}$, and this proves to be fine enough to see an effective restoration of rotational symmetry at the scale of hadrons. The temporal direction has a finer spacing corresponding to $a_t^{-1} \sim 5.6\,\mathrm{GeV}$ that gives an excellent resolution of the time-dependence of hadron two-point correlators. With the strange quark mass held fixed, light quark masses corresponding to pions of mass between 400 and 700 MeV were considered. Spatial volumes were $L^3 \sim (2.0\,\mathrm{fm})^3,  (2.5\,\mathrm{fm})^3$ and no significant trends in volume dependence were observed in the extracted spectra. This fact was used to argue that these calculations are not resolving the expected physics of excited states as \emph{resonances} in meson-meson scattering. In a finite volume one would expect a discrete and volume-dependent spectrum of meson-meson energy levels that were not observed. The spectrum that was observed was thus interpreted as `single-hadron' states whose resonant nature cannot be explored without adding to the calculation operators resembling pairs of hadrons. In this paper we will explore the possible bound-state compositions of these single-hadron states.

Quantities extracted from the lattice calculation are dimensionless, dimensionful quantities being scaled by an appropriate power of the lattice spacing - for example a mass would be extracted as $a_t m$, where, if $a_t$ is determined, we can scale-set and obtain a mass in MeV. Scale-setting in a calculation with unphysically heavy quarks is an ambiguous procedure since there is no experimental quantity that we can compare with to set the scale. We choose a simple but mass-dependent procedure to set the scale - all mass-dimension quantities are scaled as follows,
\begin{equation}
	m = \frac{(a_t m)}{(a_t m_\Omega)} m_\Omega^{\mathrm{phys.}},\nonumber
\end{equation}
where $(a_t m_\Omega)$ is the dimensionless mass of the $\Omega$ baryon computed on this lattice and $m_\Omega^{\mathrm{phys.}}$ is the experimental mass.

\begin{figure*}
 \centering
\includegraphics[width=0.8\textwidth, bb=0 30 576 360]{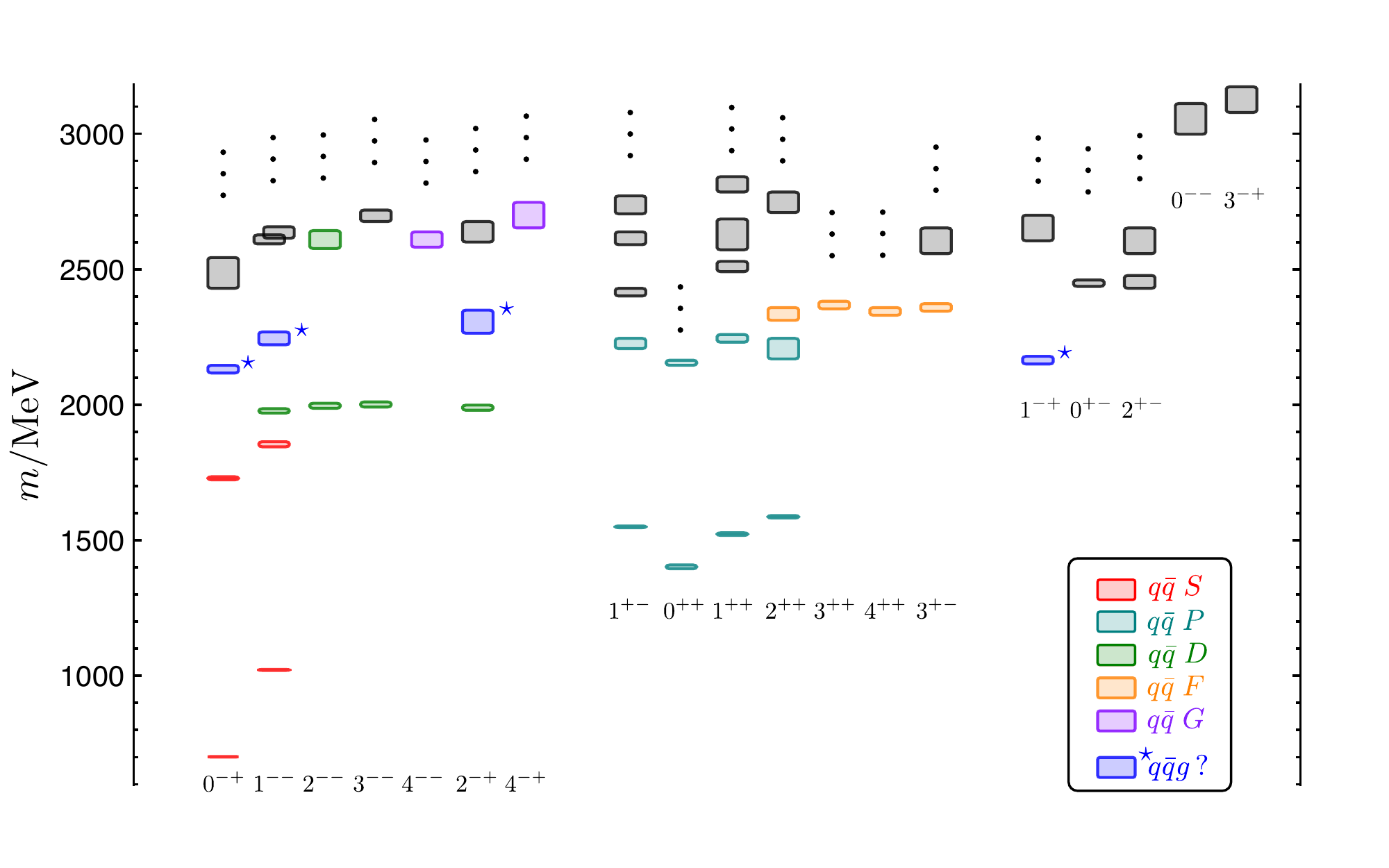}
\caption{Spectrum of isovector mesons extracted in \cite{Dudek:2010wm} with a pion mass of $\sim 700\,\mathrm{MeV}$. Masses (in MeV) are averages over two computed volumes, $\sim (2.0\,\mathrm{fm})^3, (2.5\,\mathrm{fm})^3$. Box size indicates the statistical uncertainty on the extracted mass. Ellipses indicate that there may be heavier states in that $J^{PC}$ but we did not determine them reliably in \cite{Dudek:2010wm}. 
States are color coded by assigned $L$-wave supermultiplet as described in the text. States colored grey could not be assigned to an $L$-wave supermultiplet.\label{743ass}}
\end{figure*}

We will focus initially in this paper on the spectrum results obtained at the heaviest `light' quark mass, where all three quarks are at the strange quark mass and the corresponding pion mass is $\sim 700\,\mathrm{MeV}$. We will see that the qualitative features observed in the spectrum will change relatively little as the pion mass is reduced to $\sim 400 \,\mathrm{MeV}$. The spectrum of isovector states extracted is shown in Figure 
\ref{743ass} where we see the benefits of using a large basis in the number of excited states extracted in a given $J^{PC}$ and the range of $J^{PC}$ considered.

\section{Interpretation of the meson spectrum}

\begin{table}
\begin{tabular}{|cc|c|}
\hline
\multicolumn{2}{|c|}{$q\bar{q}$}& $J^{PC}$ \\
\hline
\hline
\multirow{2}{*}{$L=0(S)$}	 & $S=0$ & $0^{-+}$ \\
						 & $S=1$ & $1^{--}$ \\
						 \hline
\multirow{2}{*}{$L=1(P)$}	 & $S=0$ & $1^{+-}$ \\						
						 & $S=1$ & $(0,1,2)^{++}$ \\ 
\hline
\multirow{2}{*}{$L=2(D)$}	 & $S=0$ & $2^{-+}$ \\						
						 & $S=1$ & $(1,2,3)^{--}$ \\ 
\hline
\multirow{2}{*}{$L=3(F)$}	 & $S=0$ & $3^{+-}$ \\						
						 & $S=1$ & $(2,3,4)^{++}$ \\ 
\hline
\multirow{2}{*}{$L=4(G)$}	 & $S=0$ & $4^{-+}$ \\						
						 & $S=1$ & $(3,4,5)^{--}$ \\ 
\hline
\end{tabular}
\caption{$q\bar{q}$ $\,^{2S+1}\!L_J$, $J^{PC}$ supermultiplets.  \label{qq}}
\end{table}

Of immediate importance is the presence in Figure \ref{743ass} of a spectrum of exotic $J^{PC}$ mesons: a lightest $1^{-+}$, significantly lighter than positive parity states, $0^{+-}, (2^{+-})^2$. As well as more $1^{-+}$ states, \emph{much} higher in the spectrum are $0^{--}$ and $3^{-+}$ states. We will argue that these exotic states can be explained as being hybrid mesons.

Under examination, the spectrum of \emph{non-exotic} $J^{PC}$ mesons shown in Figure \ref{743ass} displays $J^{PC}$ patterns consistent with $q\bar{q}\,^{2S+1}\!L_J$ structure (as presented in Table \ref{qq}). These patterns are identified in Figure \ref{743ass}, where we show apparently complete $S,P,D,F$-wave supermultiplets\footnote{The terminology ``\emph{super}multiplet" is a historical one, where flavor multiplets of a given $J^{PC}$ were first identified and later different $J^{PC}$ having a common proposed bound-state structure were related.}
, and a partial $G$-wave supermultiplet indicated by the presence of $4^{--}, 4^{-+}$ states. There appear to be higher mass recurrences of several of these supermultiplets, which in analysis to be described later we will suggest could be radial excitations of $q\bar{q}$.

Once these assignments based upon degeneracy patterns are made, we can see a near-degenerate set of states with $J^{PC} = 0^{-+}, 1^{--}, 2^{-+}$ remains unassigned (colored blue and indicated with a star ($\color{blue}\star$) in Figure \ref{743ass}). No supermultiplet in Table \ref{qq} contains such a set, and we notice that the mass-scale of these states is very close to the lightest exotic state, the $1^{-+}$. We hypothesize that these states could be members of the lightest hybrid meson supermultiplet.

To explore the hypothesis that we have a supermultiplet of hybrid mesons embedded within a spectrum of $q\bar{q}$ states, we take advantage of the ``overlap" information provided by the two-point correlator analysis described in the previous section. In the same variational analysis that determines the mass spectrum, $m_\mathfrak{n}$, we also determine the matrix-elements, $Z_i^\mathfrak{n} = \big\langle \mathfrak{n} \big| {\cal O}_i \big| 0 \big\rangle$, that encode the degree to which operator ${\cal O}_i$ ``overlaps with" state $\mathfrak{n}$. To the extent that we understand the structure of our operators, we can infer some information about the internal structure of the extracted states from the relative size of these overlaps.
 
\subsection{Operator interpretation}
As an example, consider the simplest $1^{--}$ operator included in our basis, $\bar{\psi} \gamma_i \psi$, which as a shorthand notation we call ``$\rho$". If we argue that the smeared fermion fields are capable of producing a single constituent $q\bar{q}$ pair, then writing out the quark-spin and momentum structure of the spinor contraction, $\bar{u}_\sigma(\vec{p}_q) \gamma_i v_{\bar{\sigma}}(\vec{p}_{\bar{q}})$, we realise that this operator will dominantly overlap with $q\bar{q}\, \,^3\!S_1$ states. Overlap with $q\bar{q}\, \,^3\!D_1$ comes from lower components of the Dirac spinors and is suppressed by powers of $\tfrac{|\vec{p}_q|}{m_q}$. This ``sub-leading" overlap can be further suppressed by considering the upper-component projected operator $\bar{\psi} \gamma_i \tfrac{1}{2}(1 - \gamma_0) \psi$, which we call ``$\rho_\mathrm{NR}$". 

We must emphasise here that since the fermion fields are smeared over space we are not measuring the decay constant of a vector meson with the ``$\rho$" operator. Rather we are sampling a smearing function weighted integral of the state's wavefunction \cite{Dudek:2008sz}. A bound-state model description of this can be found in Appendix A.

To extend our interpretative framework consider the simplest operator featuring a gauge-covariant derivative, $\bar{\psi} \overleftrightarrow{D}_i \psi$, which we call $\big( a_0 \times D^{[1]}_{J=1} \big)^{J=1}$ and which transforms like $J^{PC}=1^{--}$. Initially let us simply neglect the gauge-field term in the gauge-covariant derivative in our interpretation and use the ordinary derivative acting on the quark field to introduce an extra factor of the quark momentum. The resulting spinor contraction has overlaps of comparable size for both $q\bar{q}\, \,^3\!S_1$ and $q\bar{q}\, \,^3\!D_1$.

At the two derivative level, we have an example of an operator which has leading overlap with $q\bar{q}\, \,^3\!D_1$ and suppressed $q\bar{q}\, \,^3\!S_1$, namely $\big(\rho_\mathrm{NR}\times D^{[2]}_{J=2}\big)^{J=1}$. Interpreting the ordinary derivative part of the operator as giving rise to a factor $Y_2^m(\partial)$ we see where the $D$-wave structure enters. 

The same basic procedure of neglecting (for interpretation purposes) the gauge-field part of gauge-covariant derivatives can be followed for all our operators, yielding $q\bar{q}\, \,^{2S+1}\!L_J$ constructions for non-exotic $J^{PC}$, except that once we reach two-derivative constructions we hit a non-trivial case. One possible combination is $D^{[2]}_{J=1} \equiv \big\langle 1,m_1; 1,m_2 \big| 1, m\big\rangle \overleftrightarrow{D}_{m_1}\overleftrightarrow{D}_{m_2}$, where, since the Clebsch-Gordan coefficient is antisymmetric under exchange of $(m_1, m_2)$, in the limit of ignoring the gauge-field part of the derivatives we have zero. This operator construction is only non-zero precisely because the gauge-field is present and in fact is proportional to chromomagnetic components of the field-strength tensor, $F$. Since one must have a non-trivial gauge-field structure to have a non-zero value of $F$ we propose that \emph{overlap with this operator indicates hybrid-like character}.

As a concrete example of such an operator we have $\big(\rho_\mathrm{NR} \times D^{[2]}_{J=1}\big)^{J=1}$ which is an operator with exotic $J^{PC}=1^{-+}$. The structure of this operator would appear to be that of a $q\bar{q}$ pair with $S=1$ and $L=0$ (from $\rho_\mathrm{NR}$) along with a non-trivial gluonic field having chromomagnetic character ($J_g^{P_gC_g} = 1^{+-}$, from $D^{[2]}_{J=1}$). This is the kind of hybrid meson construction proposed in the bag model \cite{Barnes:1982tx, Chanowitz:1982qj} and in the ``$P$-wave" quasi-gluon picture \cite{Guo:2008yz}. In the lattice calculation we find that the lightest extracted $1^{-+}$ state has large overlap onto this operator.

\begin{figure}
 \centering
\includegraphics[width=.40\textwidth]{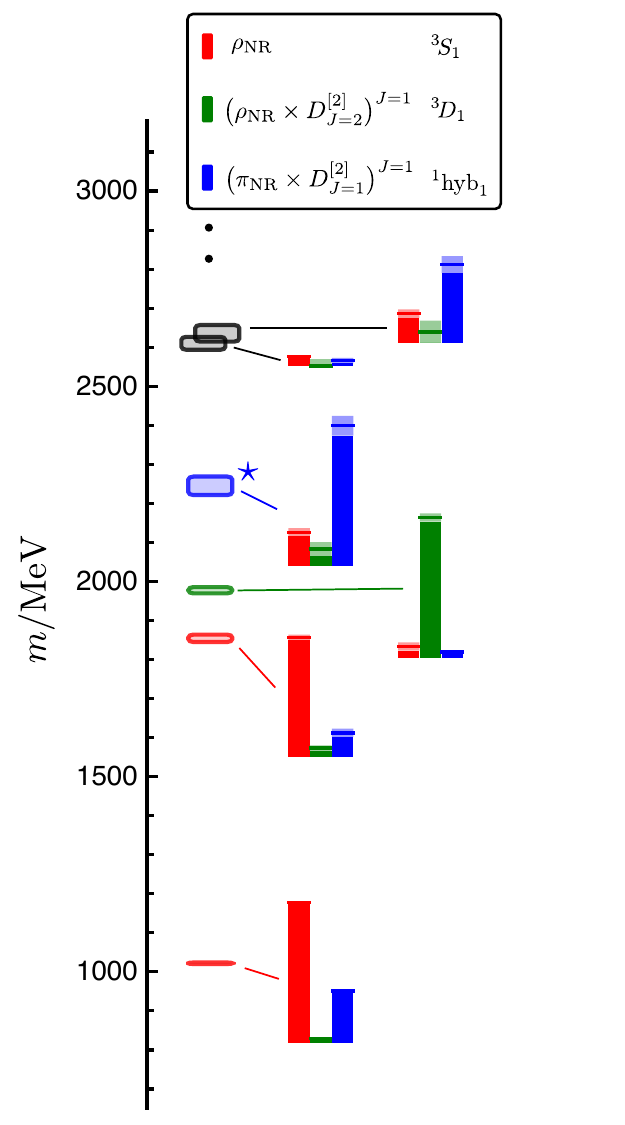}
\caption{Spectrum of $1^{--}$ mesons (as in Figure \ref{743ass}) along with relative size of overlaps onto operators, $\rho_\mathrm{NR} \sim \,^3\!S_1$, $\rho_\mathrm{NR}\times D^{[2]}_{J=2} \sim \,^3\!D_1$ and $\pi_\mathrm{NR}\times D^{[2]}_{J=1} \sim \,^1\mathrm{hyb}_1$. Histograms normalised such that the state with largest overlap across the entire excited $1^{--}$ spectrum has mean value of $1$. The lighter colored area at the head of each bar represents the statistical uncertainty.\label{histo}}
\end{figure}

\begin{figure*}
 \centering
\includegraphics[width=0.95\textwidth,  bb= 40 30 700 216]{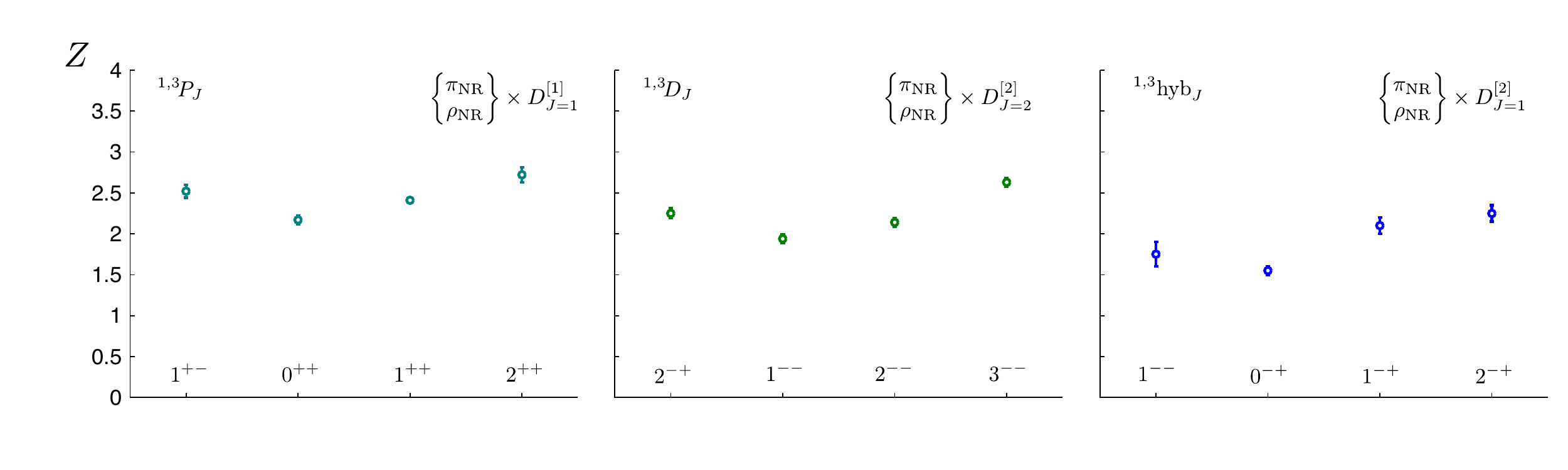}
\caption{$Z$ values in lattice units. Left pane: lightest $P$-wave supermultiplet, $1^{+-},(0,1,2)^{++}$ overlaps onto $\{\pi,\rho\}_\mathrm{NR}\times D^{[1]}_{J=1}$. Middle pane: lightest $D$-wave supermultiplet, $2^{-+},(1,2,3)^{--}$ overlaps onto $\{\pi,\rho\}_\mathrm{NR}\times D^{[2]}_{J=2}$. Right pane: lightest hybrid multiplet, $1^{--}, (0,1,2)^{-+}$ overlaps onto $\{\pi,\rho\}_\mathrm{NR}\times D^{[2]}_{J=1}$.
 \label{Z743}}
\end{figure*}

Now let us return to the spectrum of non-exotic $J^{PC}$ states. We can ask how well  the assignments of $L$-wave supermultiplets made on the basis of mass degeneracy patterns match with the overlaps onto operators having dominant $q\bar{q}\, \,^{2S+1}\!L_J$ character? As an example, in Figure \ref{histo}, we present the $1^{--}$ spectrum and the overlaps onto three characteristic operators: $\rho_\mathrm{NR}$, $\big(\rho_\mathrm{NR} \times D^{[2]}_{J=2}  \big)^{J=1}$ and $\big(\pi_\mathrm{NR} \times D^{[2]}_{J=1}  \big)^{J=1}$. These three operators are dominantly $q\bar{q}\, \,^3\!S_1$, $q\bar{q}\, \,^3\!D_1$ and, according to the proposal above, hybrid with quarks in a spin-singlet. The ground state we observe to be dominantly $q\bar{q}\, \,^3\!S_1$, and similarly the first excited state, which we interpret as a radial excitation, having the same angular and spin structure as the ground state\footnote{We should point out here that a radial excitation can have an `accidentally' small overlap with an operator whose angular dependence accurately characterises the state's structure - see Appendix A for an explicit demonstration}. The second excited state, which we earlier assigned to a $D$-wave supermultiplet indeed is dominated by $q\bar{q}\, \,^3\!D_1$. It is notable that there appears to be relatively little mixing between $q\bar{q}\, \,^3\!S_1$, $q\bar{q}\, \,^3\!D_1$ basis states, even though the resulting eigenstates are very close in mass. The third excited state is the $1^{--}$ state that could not be placed into an $L$-wave supermultiplet on the basis of degeneracy patterns. We find that it has dominant overlap with the operator $\big(\pi_\mathrm{NR} \times D^{[2]}_{J=1}  \big)^{J=1}$ that we have characterised as being a spin-singlet hybrid meson.

\begin{figure}
 \centering
\includegraphics[width=0.5\textwidth, bb= 40 20 420 280]{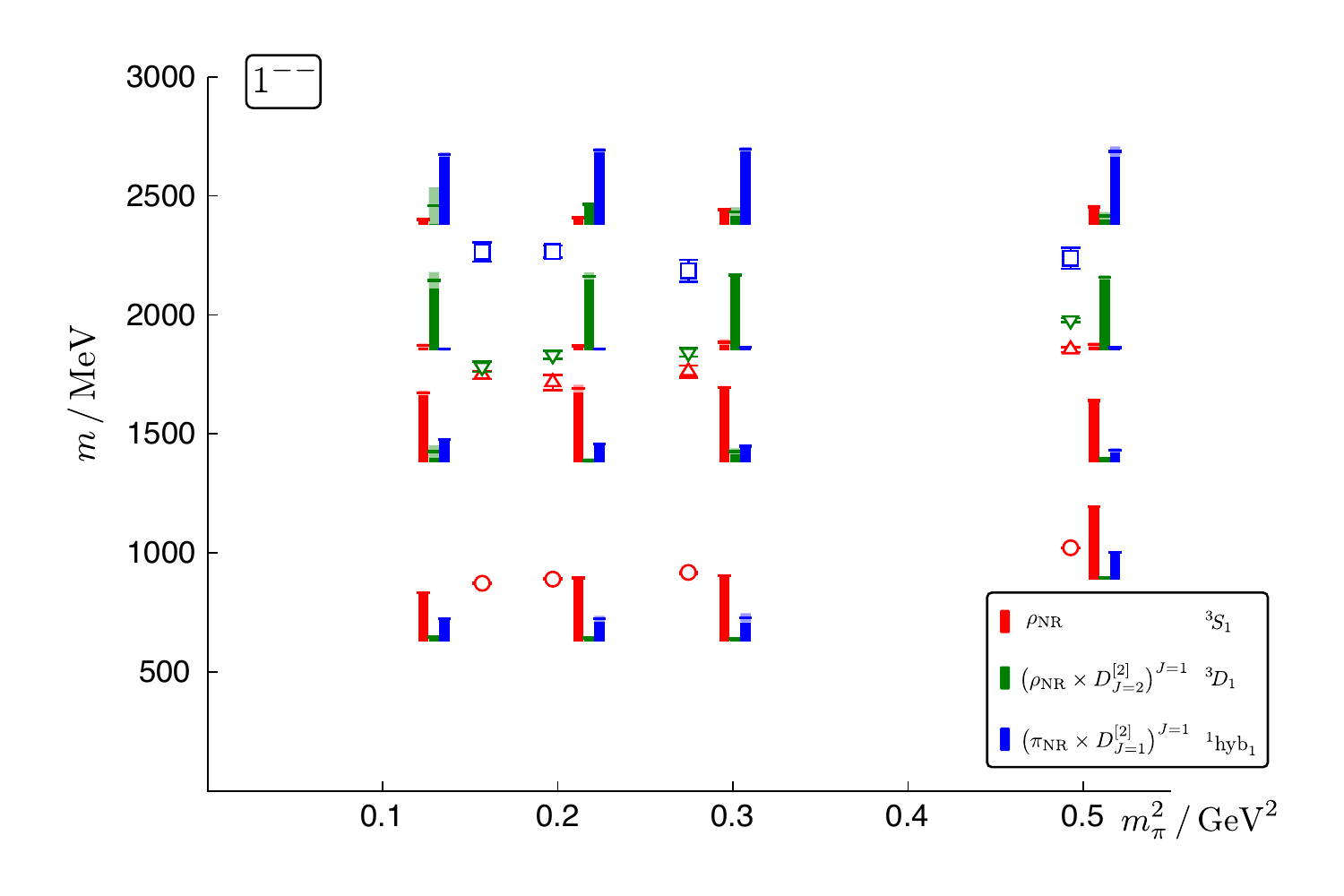}
\includegraphics[width=0.5\textwidth, bb= 40 20 420 280]{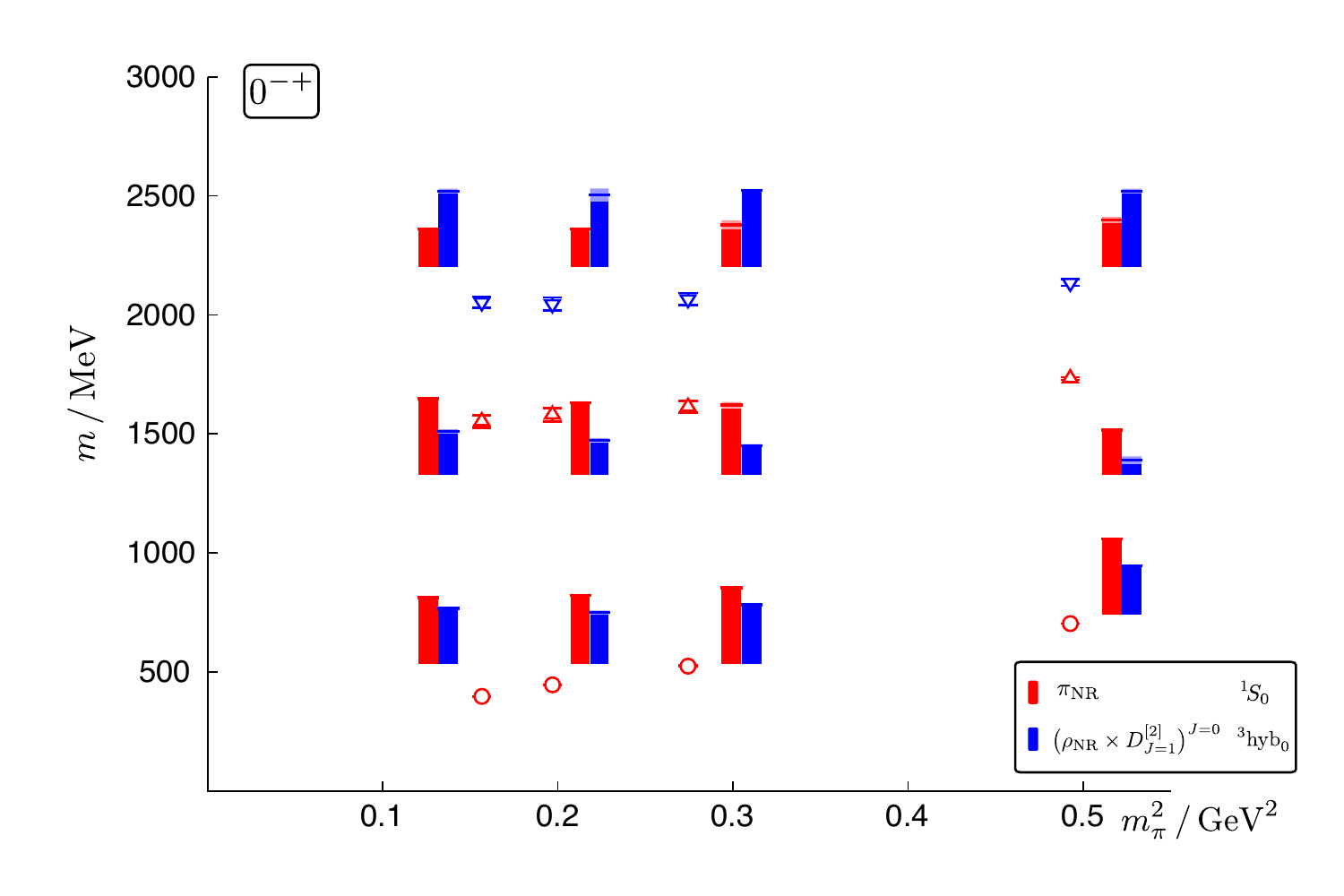}
\includegraphics[width=0.5\textwidth, bb= 40 20 420 280]{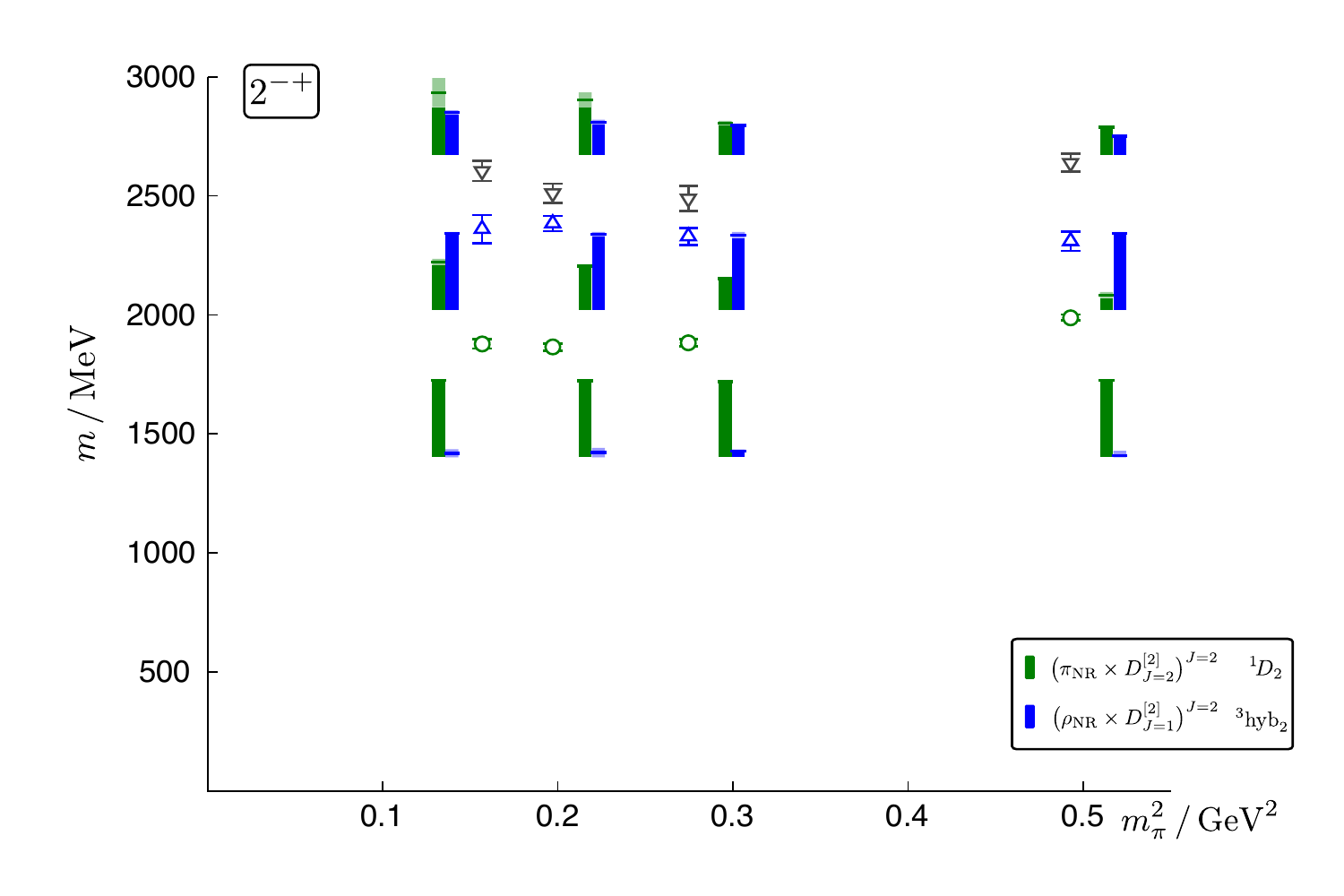}
\caption{Mass spectrum and histograms as in Figure \ref{histo} as a function of quark mass expressed via the pion mass. \label{massplots}}
\end{figure}

Very similar conclusions can be drawn in the $2^{-+}$ channel\footnote{see Figure \ref{massplots}}, where the ground state is found to be dominantly $q\bar{q}\, \,^1\!D_2$, while the first excited state overlaps strongly onto $\big(\rho_\mathrm{NR} \times D^{[2]}_{J=1}  \big)^{J=2}$ and which we thus propose is a hybrid meson with quarks in a spin-triplet.

The $0^{-+}$ channel is not quite so clearly delineated - the ground state has a large overlap onto operators of $q\bar{q}\, \,^1\!S_0$ character, but also onto a hybrid $\big(\rho_\mathrm{NR} \times D^{[2]}_{J=1}  \big)^{J=0}$ operator. That the pion might have behavior not entirely compatible with constituent $q\bar{q}$ structure is to be expected given its role as the pseudo Goldstone boson of spontaneous chiral symmetry breaking, and we will not explore it in detail here. The first excited state appears more conventional, with overlap onto $q\bar{q}\, \,^1\!S_0$ operators and suppressed overlap onto $\big(\rho_\mathrm{NR} \times D^{[2]}_{J=1}  \big)^{J=0}$. The second excited state has the largest $\big(\rho_\mathrm{NR} \times D^{[2]}_{J=1}  \big)^{J=0}$ overlap, but unlike the $1^{--}, 2^{-+}$ hybrid candidates, it also has significant overlap onto $q\bar{q}$ operators. This may indicate a larger degree of mixing with $q\bar{q}$ basis states than is present in the other non-exotic channels present in the supermultiplet.

In identifying a supermultiplet, one expects the $Z$ values from a common operator projected across the allowed $J^{PC}$ to be similar. For example, $\rho_\mathrm{NR} \times D^{[1]}_{J=1}$ can be projected into $J^{PC}=(0,1,2)^{++}$ using appropriate Clebsch-Gordans and the corresponding spin-singlet comes from $\pi_\mathrm{NR} \times D^{[1]}_{J=1}$. The normalisation is such that one would expect the $Z$ value extracted to be common to all members of the supermultiplet. We show the extracted $Z$ values for the lightest proposed $q\bar{q}$ $P$-wave supermultiplet in the left-hand pane of Figure \ref{Z743} where we see reasonable agreement, supporting the assignment. In the middle pane of Figure \ref{Z743} we show a similar pattern for the lightest proposed $q\bar{q}$ $D$-wave supermultiplet. The right-hand pane shows the pattern for our proposed lightest hybrid multiplet using the operators $\big(\{\pi,\rho\}_\mathrm{NR} \times D^{[2]}_{J=1}  \big)^{J}$, that strongly suggests these four states have a common structure\footnote{A model-dependent demonstration using a constituent gluon construction is given in Appendix B.}.

We caution the reader at this point that only qualitative judgements can be drawn from the value of the $Z$s since these matrix-elements determined with lattice regularisation require renormalisation to be compared to the continuum theory matrix-elements. That renormalisation can mix operators, including those of differing mass dimension, although with a relatively fine lattice spacing, an improved action, and smeared fields we do not anticipate such mixing being large.

We have now demonstrated the main result of this paper, the first identification of the lightest hybrid supermultiplet within QCD. 
As well as a single exotic $1^{-+}$ state, there are three non-exotic quantum numbered states, $0^{-+}, 1^{--}, 2^{-+}$, that differ from the corresponding $q\bar{q}$ constructions by having the opposite quark-spin (singlet versus triplet).

\begin{figure*}
 \centering
\includegraphics[width=0.95\textwidth
]{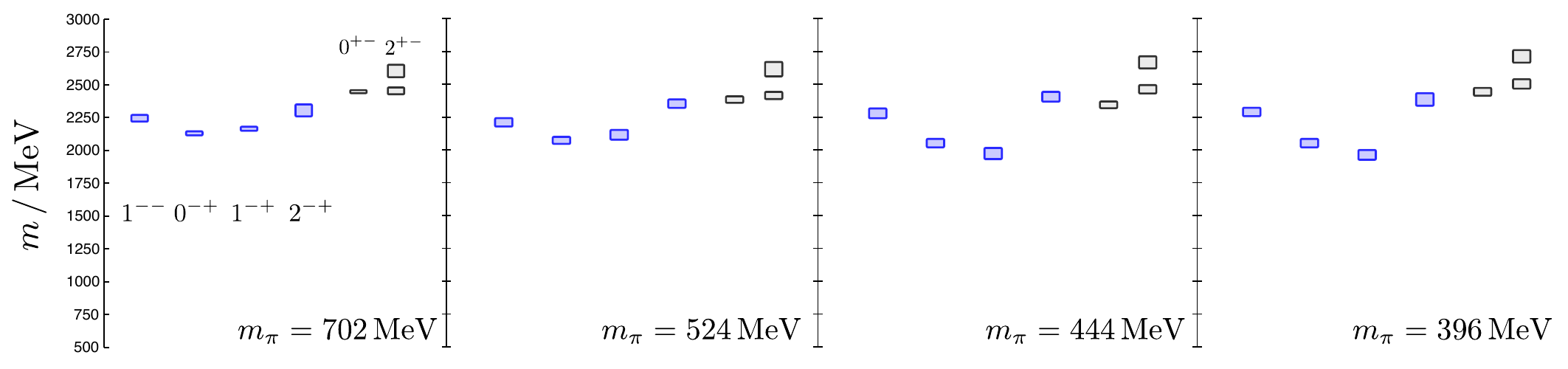}
\caption{Mass spectrum of lightest isovector hybrid supermultiplet along with lightest positive parity exotic states computed at four pion masses.\label{hyb}}
\end{figure*}

The reader may be concerned that the results presented thus far are for a version of QCD in which the three `light' quark flavors are all at the strange quark mass, and this world may have little to do with the version of QCD manifested in this universe which has two flavors being much lighter than strange. The possibility of significant changes under reduction of the light quark mass can be explored in lattice QCD calculations with lighter quark masses and calculations down to $m_\pi \sim 400 \,\mathrm{MeV}$ were reported in \cite{Dudek:2010wm}. The same analysis in terms of $q\bar{q}$ $L$-wave supermultiplet degeneracies and operator overlaps has been carried out and the qualitative features remain unchanged. Relevant to the particularly important case of the lightest hybrid supermultiplet, we show in Figure \ref{massplots} the quark mass dependence of the $1^{--},0^{-+}$ and $2^{-+}$ spectra and operator overlaps. We see that qualitatively there is relatively little change, except possibly in the degree of $q\bar{q}$-hybrid mixing in the $(0,2)^{-+}$ channels. The mass splitting within the supermultiplet grows with decreasing quark mass as shown in Figure \ref{hyb}, but the lightest negative parity hybrid states remain systematically below the positive parity exotics and we argue that it remains sensible to describe them as a supermultiplet\footnote{the large mass of the $2^{-+}$ state may be due to significant mixing with a heavier $q\bar{q}\,^1\!D_2$ basis state as suggested by the overlaps in Figure \ref{massplots}.}. 

In the remainder of the paper we will be concerned with extending our description of the spectrum, identifying higher mass hybrid states, comparing with models of gluonic excitations and finally with developing a phenomenology of hybrid mesons that is directly motivated by results of QCD calculations.

\subsection{Heavier hybrid mesons}

Having identified what we propose is the lightest supermultiplet of hybrid mesons, we now move on to consider the presence of heavier hybrid meson states in the extracted lattice QCD spectrum. The first obvious candidates in the $m_\pi \sim 700 \,\mathrm{MeV}$ calculation are the exotic $0^{+-}$ and $2^{+-}$ states above 2.4 GeV. The $0^{+-}$ state is found to have considerable overlap with the $\big( a_1 \times D^{[2]}_{J=1} \big)^{J=0}$ operator, which has the same gluonic structure ($D^{[2]}_{J=1}$) as the hybrids previously identified, but which places the spin-triplet $q\bar{q}$ pair in a relative $P$-wave ($a_1 \sim \bar{\psi} \gamma_5 \gamma_i \psi \sim q\bar{q}\,^3\!P_1$). This suggests that there might be an excited hybrid supermultiplet with the same chromomagnetic gluonic excitation, but also excited in quark angular momentum to $L_{q\bar{q}}=1$. Such a supermultiplet would be expected to house states with $J^{PC} = \{ (0,1,2)^{++}, 1^{+-}\}_{q\bar{q}} \otimes 1^{+-}_g = \mathbf{0^{+-}}, (1^{+-})^3, (\mathbf{2^{+-}})^2, 3^{+-}, 0^{++}, 1^{++}, 2^{++}$ where the exotic states are shown in bold. The presence of a single $0^{+-}$ and two $2^{+-}$ states is indeed observed in the spectrum with relatively little splitting between them. In order to find non-exotic members of this supermultiplet embedded within the $q\bar{q}$ states one should examine the overlaps for a wider set of operators than we have so far considered including those constructed from three derivatives. We will not go into that level of detail in this paper.

There remain further exotic states at higher mass. The heavier $1^{-+}$ state at about 2.6 GeV has a pattern of overlaps that is similar to the lightest $1^{-+}$ and it thus may correspond to a `radial excitation' in which the quark radial motion is excited on top of the same chromomagnetic gluonic excitation. The heaviest $1^{--}$ state shown in Figure \ref{histo}, which has a significant overlap onto $\pi_\mathrm{NR}\times D^{[2]}_{J=1}$, could conceivably be another member of a radially excited hybrid supermultiplet - the remaining $(0,2)^{-+}$ members are not identified in this calculation. 

The $3^{-+}$ state above 3 GeV could be part of a supermultiplet having $q\bar{q}$ in $D$-wave coupled to the lowest chromomagnetic gluonic excitation, or it could arise from a different gluonic excitation. One possibility would be $S$-wave $q\bar{q}$ with gluonic field transforming as $J^{P_gC_g}_g = 2^{+-}$. At the level of detail we have in this calculation we cannot definitively state which is the case. 

There is an exotic $0^{--}$ state that appears at a very heavy mass above 3 GeV - such a state does not arise with a chromomagnetic gluonic excitation combined with $q\bar{q}$ and most likely signals the mass scale for a different type of gluonic excitation, possibly one having $J^{P_gC_g}_g = 1^{--}$. 

We note here that the energy ordering of $J_g^{P_gC_g}$ as $1^{+-}\!<\!1^{--}\!<\!\ldots$ is the same as is observed for gluelumps, the excitations of the gluonic field around a color-octet source in $SU(3)$ Yang-Mills theory\cite{Bali:2003jq}.

\subsection{Isoscalar mesons and kaons}

So far we have identified $J^{PC}$ \emph{super}multiplets within the isovector meson spectrum but have not considered flavour partners in a flavour multiplet. We expect there to be kaonic and isoscalar states within flavor octets and singlets associated with the isovector states. Kaons have a denser spectrum described by the more limited $J^P$ quantum numbers, the flavored states not being eigenstates of charge-conjugation (or the $G$-parity extension). An immediate consequence of this is that there are no exotic kaons, and hence no smoking gun signature for hybrid kaons.
Within simple $q\bar{q}$ models, the spectrum of kaons features states constructed from admixtures of opposite $C$, for example in the axial ($1^+$) sector one would expect two low-lying states constructed from the basis states $\bar{u}s\,^3\!P_1(1^{+(+)})$, $\bar{u}s\,^1\!P_1(1^{+(-)})$. Experimentally two states are found, the $K_1(1270), K_1(1400)$, whose decay properties suggest that they are strong admixtures of the above basis states, with a mixing angle close to $45^\circ$\cite{Nakamura:2010zzi}. 

Kaon spectra were extracted in the lattice calculations reported in \cite{Dudek:2010wm}  using the derivative basis described earlier, but using both $J^{P+}$ and $J^{P-}$ operators together in the $J^P$ calculation. As such, by considering overlaps $\big\langle \mathfrak{n}(J^P) \big| {\cal O}(J^{P(C)}) \big| 0 \big\rangle$ one can infer the degree of mixing between $C=+$ and $C=-$ basis states. In explicit computation we found this mixing to be very small, but this is likely to be a reflection of the unphysically heavy light quark masses used (the lightest is $m_\pi \sim 400 \,\mathrm{MeV}$) which do not give rise to as large a breaking of $SU(3)_F$ as is present in nature. In light of this, and the impossibility of exotic quantum numbered states, we will not discuss kaons any further.

\begin{figure*}
 \centering
\includegraphics[width=0.99\textwidth
]{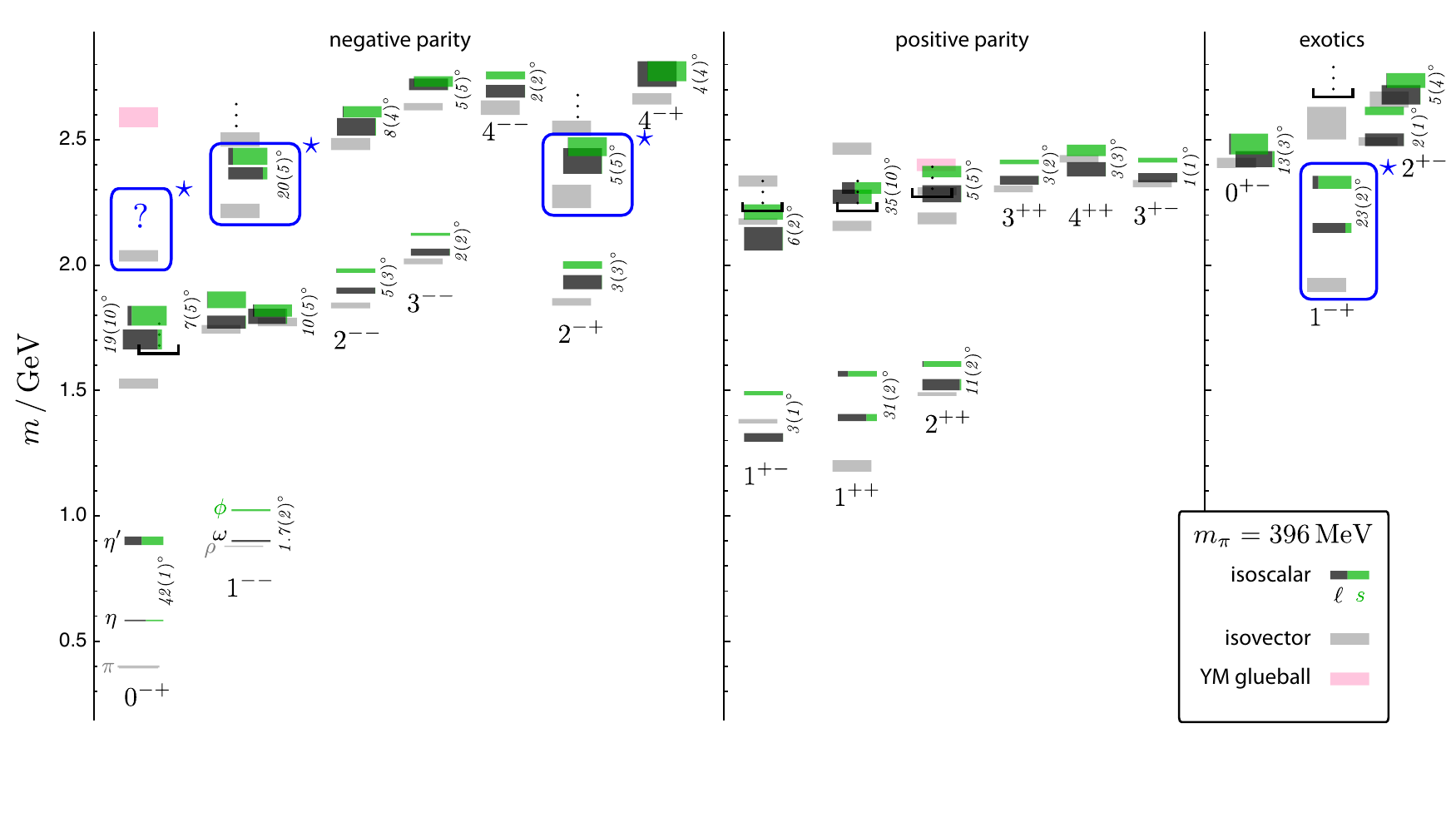}
\caption{Isoscalar meson spectrum with $m_\pi \sim 400\,\mathrm{MeV}$ labeled by $J^{PC}$ (reproduced from \cite{Dudek:2011tt}).  The light-strange content of each state ($\cos^2 \alpha, \sin^2 \alpha$) is given by the fraction of (black, green) and the mixing angle for identified pairs is also shown. Horizontal square braces with ellipses indicate that additional states were extracted in this $J^{PC}$ but were not robust.  Grey boxes indicate the positions of isovector meson states extracted on the same lattice (taken from \cite{Dudek:2010wm}). Pink boxes indicate the position of glueballs in the quark-less Yang-Mills theory \cite{Morningstar:1999rf}. The candidate states for the lightest hybrid meson supermultiplet are indicated by the blue boxes and stars.\label{iso}}
\end{figure*}

Isoscalar meson properties are somewhat more challenging to compute in lattice QCD owing to the need to evaluate disconnected Wick contractions. Distillation \cite{Peardon:2009gh} proves to be an efficient procedure to do this and in \cite{Dudek:2011tt} we reported a calculation of the isoscalar meson spectrum at a single pion mass, $m_\pi \sim 400 \,\mathrm{MeV}$ using the operator basis of \cite{Dudek:2010wm}, doubled in size by including both light-quark ($\ell \equiv \tfrac{1}{\sqrt{2}}\big(\bar{u} \mathbf{\Gamma} u + \bar{d} \mathbf{\Gamma} d  \big)$) and strange-quark bilinears. The spectrum obtained in a $(2.0\,\mathrm{fm})^3$ box is reproduced in Figure \ref{iso}. The degree to which a given state is dominated by $\ell$ or $s$ can be estimated using the relative size of overlaps onto operators. We parameterize the admixture by introducing a mixing angle, $\alpha$, shown in Figure \ref{iso}, assuming pairs of states ($\mathfrak{a}$, $\mathfrak{b}$) are orthogonal combinations of just two light-strange basis states: 
\begin{align}
\big| \mathfrak{a} \big\rangle &= \cos \alpha \big| \ell \big\rangle - \sin \alpha \big| s\big\rangle \nonumber\\
\big| \mathfrak{b} \big\rangle &=  \sin \alpha \big| \ell \big\rangle + \cos \alpha \big| s\big\rangle. \nonumber
\end{align}

We note that the basic qualitative phenomenology observed in experiment is reproduced. In most $J^{PC}$ the flavor mixing is close to ideal, giving rise to almost pure $\tfrac{1}{\sqrt{2}}\big(u\bar{u} + d\bar{d}  \big)$ and $s\bar{s}$ states. The exceptions to this are the $0^{-+}$ channel where the $\eta,\eta'$ mixing being quite close to $SU(3)_F$ ideal octet-singlet is observed and the $1^{++}$ channel where phenomenological analysis of experimental radiative transition rates suggest the $f_1(1285), f_1(1420)$ are somewhat flavor mixed as seen in the lattice data. 

Of particular interest to us here is the spectrum of exotic isoscalar mesons, where the basic structure of $1^{-+}$ lightest below $0^{+-}$ and $2^{+-}$ observed for isovectors is again seen, but doubled to reflect the presence of both light and strange quarks. The positive parity exotics are very close to being ideally flavor mixed, but the $1^{-+}$, which in the isovector case we placed in a different supermultiplet, has a noticable degree of light-strange mixing. The large amount of strange-quark in the lighter state is most likely the dominant reason for it being significantly heavier than the corresponding isovector state\footnote{Note that suggestions from $SU(3)$ Yang-Mills calculations are that the lightest exotic $J^{PC}$ glueball is much heavier than these states\cite{Morningstar:1999rf}, so we are justified in proceeding with a hybrid meson interpretation.}.

\begin{figure}
 \centering
\includegraphics[width=0.40\textwidth]{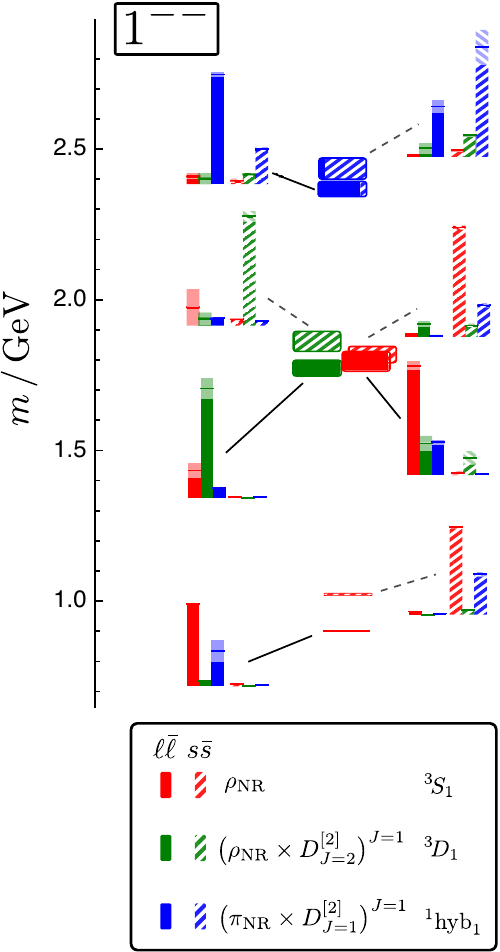}
\caption{Spectrum of $1^{--}$ isoscalar mesons, as in Figure \ref{iso} along with relative size of overlaps onto operators, $\rho_\mathrm{NR} \sim \,^3\!S_1$, $\rho_\mathrm{NR}\times D^{[2]}_{J=2} \sim \,^3\!D_1$ and $\pi_\mathrm{NR}\times D^{[2]}_{J=1} \sim \,^1\mathrm{hyb}_1$. Solid bars are for light-quark overlap and dashed bars for strange. \label{isohisto}}
\end{figure}

In Figure \ref{isohisto} we show overlap histograms for the isoscalar $1^{--}$ states. We see that the $\,^3\!S_1 < \,^3\!S_1 < \,^3\!D_1 < \,^1\mathrm{hyb}_1$ pattern is again present, but doubled for light and strange quarks. The hybrid states show a greater degree of light-strange flavor mixing than do the $q\bar{q}$ states. Similarly we identify that the $2^{-+}$ states near 2.4 GeV have strong overlap with the $\rho_\mathrm{NR} \times D^{[2]}_{J=1}$ operators. Signals in the $0^{-+}$ sector are not clear enough to identify an anticipated pair of hybrid states above 2 GeV that would complete the hybrid supermultiplet.

In summary we may well be observing the same hybrid supermultiplet structure in the isoscalar sector, and there appears to be some significant flavor mixing away from pure $\tfrac{1}{\sqrt{2}}\big(u\bar{u} + d\bar{d}  \big)$, $s\bar{s}$ structure.


\section{Comparison with models}

The models described earlier can now be examined in light of the hybrid spectrum extracted from an interpretation of lattice QCD. We begin with the flux-tube model which predicts a large lightest supermultiplet: $(0,1,2)^{-+}, 1^{--}, (0,1,2)^{+-}, 1^{++}$. The first four of these states have been clearly identified in the lattice QCD isovector spectrum, and one might argue that there are candidates for the remaining four at slightly higher energy. There would need to be considerable splitting within the supermultiplet and since \emph{two} $2^{+-}$ states appear to be present in the mass region under consideration, there would need to be a further flux-tube excitation almost overlapping with the first. This appears to be an inefficient modeling of the observed spectrum. 

Bag models having a lowest energy ``TE" gluon mode are chromomagnetic in gluonic character and give rise to a lightest hybrid supermultiplet that agrees with the one we observe. In \cite{Chanowitz:1982qj} perturbative computations suggest that mixing of the non-exotic hybrid states with $q\bar{q}$ states should be small. There is some evidence in the lattice calculation (presented in Figure \ref{massplots}) for mixing for the $0^{-+}$ and $2^{-+}$ states, but we have not attempted here to quantify the degree. Heavier hybrid meson states were not studied in detail in the bag model.

Models in which a constituent gluon ($J^{P_gC_g}_g=1^{--}$) is added in $S$-wave to a $q\bar{q}$ system completely fail to describe our observed states, giving rise to non-exotic positive parity states with quarks in an $S$-wave and exotic $0^{--}$ and $1^{-+}$ with quarks in a $P$-wave. The $0^{--}$ is very high-lying in our extracted exotic spectrum and this suggests that this kind of hybrid construction does not contain the lowest energy gluonic excitation. 

If, one the other hand, we place a constituent gluon in a $P$-wave relative to the $q\bar{q}$, and couple so that $J_g^{P_gC_g}=1^{+-}$, we appear to be able to successfully describe both the lightest hybrid supermultiplet of $(0,1,2)^{-+}, 1^{--}$ (by having $q\bar{q}$ in an $S$-wave) and the heavier exotic states, $0^{+-}, (2^{+-})^2$ (with $q\bar{q}$ in a $P$-wave).	 In Appendix B we will present an explicit constituent gluon state construction that can describe many of the overlap patterns observed.

We should mention that another possible extension to the $q\bar{q}$ model of hadrons calls for the presence of states dominated by $qq\bar{q}\bar{q}$ configurations and simple constituent quark mass counting suggests that they may enter at an energy scale similar to hybrid mesons. The main evidence against the importance of such constructions is the lack of any clear signals for exotic flavor mesons. Nevertheless it is important to include interpolators with good overlap onto such quark configurations into lattice calculations and appropriate constructions are currently underway for use in future computations.

\section{Hybrid Meson Phenomenology}

In this section we wish to address the current and possible future experimental situation in light of what we have discovered about hybrid mesons in QCD. This will necessarily be somewhat speculative, as we have only determined the spectrum at unphysically heavy light quark masses, but we suspect that our qualitative observations will likely be robust under further reduction of $m_\pi$.

There is some evidence for states that might be candidates for our lightest isovector supermultiplet\cite{Nakamura:2010zzi}. An exotic $\pi_1$ near 1.6 GeV could be partnered with the claimed second excited $\pi(1800)$ and a first excited $\pi_2(1880)$ to make up the full set of spin-triplet states.

There is no clear candidate for the $\rho$-like member of the supermultiplet. This spin-singlet hybrid state would not be expected to be produced in $e^+e^-$ collisions, unless it mixes considerably with a $q\bar{q}\,^3\!S_1$ basis state. Figure \ref{massplots} suggests it may not, but we must not infer too much from that plot. It presents the overlap for a particular smearing function that may accidentally suppress the overlap of a radially-excited state - some evidence for this kind of behavior is seen in Appendix A where a range of smearing functions are evaluated. Future lattice calculations should compute directly the vector decay constants of excited states ($\bar{\psi}\gamma^\mu\psi$ overlap with unsmeared fields, suitably renormalised) to more directly determine the possibility of production in $e^+ e^-$. Other production processes such as photo- or electro-production off a hadron target may be more efficient than $e^+e^-$ and again here one can in future use a lattice calculation to aid understanding. To the extent that photo- and electro- production can be modeled by $t$-channel meson exchange, radiative transition matrix elements, $\big\langle M' \big| \bar{\psi}\gamma^\mu \psi\big| M\big\rangle$, control the rate of production. These quantities can be extracted from lattice calculations, even for excited states, although to date only the charmonium sector has been considered\cite{Dudek:2009kk}.

Recently \cite{Aubert:2006bu}, a new isoscalar $1^{--}$ state, $Y(2175)$, has been observed in $e^+ e^-$ decaying into $\phi f_0(980)$ and $\phi \eta$. This at first sight serves as a possible $s\bar{s}$ hybrid candidate, but production in $e^+e^-$ and the decay into $\phi + X$ suggest quark spin-triplet structure (with the theoretical prejudice that quark spin tends to be preserved in such a decay). This is in disagreement with the quark spin-singlet structure of hybrid vector states that we have observed in this study. One way out of this is to note that the partial decay fractions to $\phi +X$ have not been measured and may in fact make up only a small contribution to the total width. The appropriate spin-singlet decays would be $h_1^{s\bar{s}} +X$, but since almost nothing is known about the candidate state $h_1(1380)$, observing these decays is not straightforward experimentally.

One reason to propose hybrid character for a state is to argue that there is an overpopulation of states with respect to $q\bar{q}$ expectations, as has happened with the $Y(4260)$ in the charmonium region. Such an argument does not hold water here as there is only one identified $\phi$ excitation, the $\phi(1680)$, while we might expect at least two: a radially excited $s\bar{s}\,^3\!S_1$ and $s\bar{s}\,^3\!D_1$. We may have to content ourselves with a more mundane explanation of the $Y(2175)$, certainly there is currently no overwhelming reason to assume it is a hybrid meson. 

In our lattice calculation of isoscalar mesons we found that the $1^{--}$ members of the lightest hybrid supermultiplet may have a significant mixing of $\tfrac{1}{\sqrt{2}}\big( u\bar{u} + d\bar{d} \big)$ and $s\bar{s}$, in contrast to the conventional vector states which appear to be very close to ideally flavor mixed. We need to verify that this property remains true as one reduces the quark mass, but it may eventually be useful as a phenomenological filter for hybrid character in the experimental isoscalar vector meson spectrum.


\section{Summary}

We propose that, using the results of lattice QCD computations, we have identified the isovector members of lightest supermultiplet of hybrid mesons embedded within a spectrum of conventional mesons. The supermultiplet has quark spin-triplet states of $J^{PC} = (0,1,2)^{-+}$ and a quark spin-singlet state $1^{--}$. The identification followed from the presence of an exotic $J^{PC}=1^{-+}$ state, the lack of convincing conventional $q\bar{q}$ supermultiplet(s) to describe the degeneracy pattern of the other states and the common overlap of these four states onto composite QCD operators having essentially gluonic structure.

The form of the gluonic structure present in the operators having good overlap with these states is chromomagnetic, having $J_g^{P_gC_g}=1^{+-}$. With the $q\bar{q}$ pair in an internal $S$-wave this describes the observed $J^{PC}$. Heavier exotic hybrids $0^{+-}, (2^{+-})^2$ are proposed to be $q\bar{q}$ $P$-waves coupled to the same gluonic excitation. A much heavier exotic $0^{--}$ state likely indicates the next distinct gluonic excitation form which might have $J_g^{P_gC_g}=1^{--}$.

The lattice calculations interpreted to come to the above conclusions are performed with light quark masses heavier than those realised in nature. Over the limited range considered (corresponding to pion masses between 400 MeV and 700 MeV) the qualitative identification of the lightest hybrid multiplet appears to be robust. Future calculations will explore lighter quark masses.

The nonexotic $J^{PC}$ hybrid mesons will not be easily identified in experiment unless they happen to have some characteristic hadronic decays as suggested in certain models \cite{Isgur:1985vy,  Page:1998gz, Chanowitz:1982qj}. In order for a lattice QCD calculation to address the hadronic decays of hybrid mesons one requires there to be inclusion into the basis of operators some resembling pairs of mesons. Work in this direction is beginning with the mapping out of the amplitude for $\pi\pi$ scattering in the non-resonant isospin-2 \cite{Dudek:2010ew} channel and the resonant isospin-1 channel that contains the $\rho$ \cite{Feng:2010es}.


\begin{acknowledgments}
JJD gratefully acknowledges the contributions of his colleagues within the Hadron Spectrum Collaboration and useful communications with E.S. Swanson and P. Guo. Authored by Jefferson Science Associates, LLC under U.S. DOE Contract No.
DE-AC05-06OR23177. 
The U.S. Government retains a non-exclusive, paid-up,
irrevocable, world-wide license to publish or reproduce this manuscript for
U.S. Government purposes.
\end{acknowledgments}

\bibliography{biblio}

\appendix

\begin{widetext}

\section{The effect of quark-field smearing in a bound-state model}

Within a constituent quark model, we propose a decomposition of the distillation-smeared fermion field as
\begin{align}
\widetilde{\psi}^\mathsf{i}(x) = \int\!\!& \frac{d^3 \vec{k}}{(2\pi)^3}\sum_{\sigma=\uparrow\downarrow} S_q(|\vec{k}|)\left[ e^{-i k x} u_\sigma(\vec{k}) \mathsf{b}^\mathsf{i}_\sigma(\vec{k}) + e^{i k x} v_\sigma(\vec{k}) \mathsf{d}^{\mathsf{i}\dag}_\sigma(\vec{k})  \right] \label{quark}
\end{align}
where $\mathsf{b}^\mathsf{i}_\sigma(\vec{k})$ annihilates a constituent quark of color $\mathsf{i}$, spin $\sigma$ and momentum $\vec{k}$ and $\mathsf{d}^{\mathsf{i}\dag}_\sigma(\vec{k})$ creates a constituent antiquark. The transformation that turns current quarks into constituent quarks will be not be specified here, we shall only assume that it is linear - an example would be the Bogoliubov transformation. We can allow there to be non-trivial energy-momentum dispersion, which might be expressed as a `running' constituent quark mass, $m_q(|\vec{k}|)$ featuring in the spinors.

The standard implementation of distillation includes, with equal weight, a fixed number of eigenvectors of the gauge-covariant laplacian, $-\nabla^2 \xi_n = \lambda_n \xi_n$, as $\sum_n^N  \xi_n \xi_n^\dag$, which amounts to including all modes below some eigenvalue cutoff. We will include this in our model as a cutoff in quark momentum space, $S_q(|\vec{k}|) = \Theta(\Lambda - |\vec{k}|)$. For a non-interacting theory in which the eigenvectors are plane waves $\xi^\mathsf{i}(\vec{x}) \propto e^{i\vec{k}\cdot\vec{x}}\, \delta_{\mathsf{i}, \mathsf{RGB}}$ subject to cubic boundary conditions, this modeling would be essentially exact. In the interacting theory there is still a similarity between the extracted eigenvalue distribution to the momenta allowed to a free colored particle in a cubic box as seen in Figure \ref{distill}.

\begin{figure}[h]
 \centering
\includegraphics[width=0.5\textwidth
]{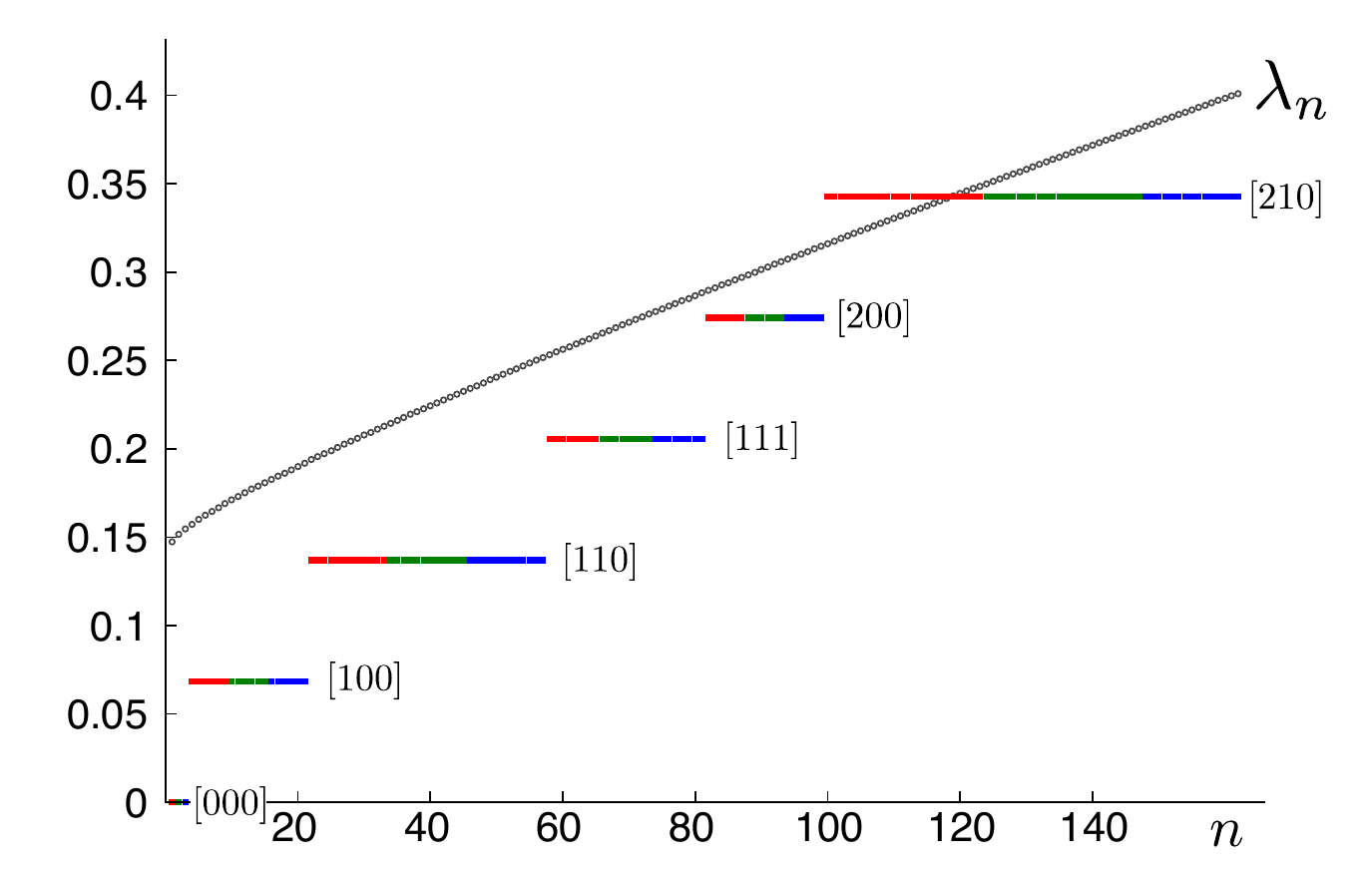}
\caption{Grey points: Eigenvalues $\lambda_n$ of $-\nabla^2$ on a $24^3$ dynamical lattice with $m_\pi \sim 400 \,\mathrm{MeV}$. Colored Boxes: $k^2 = \left(\tfrac{2\pi}{L}\right)^2(n_x^2+ n_y^2 +n_z^2)$ - allowed plane-wave momenta in a cube of side-length $L=24$.  \label{distill}}
\end{figure}

\begin{figure}
 \centering
\includegraphics[width=0.45\textwidth
]{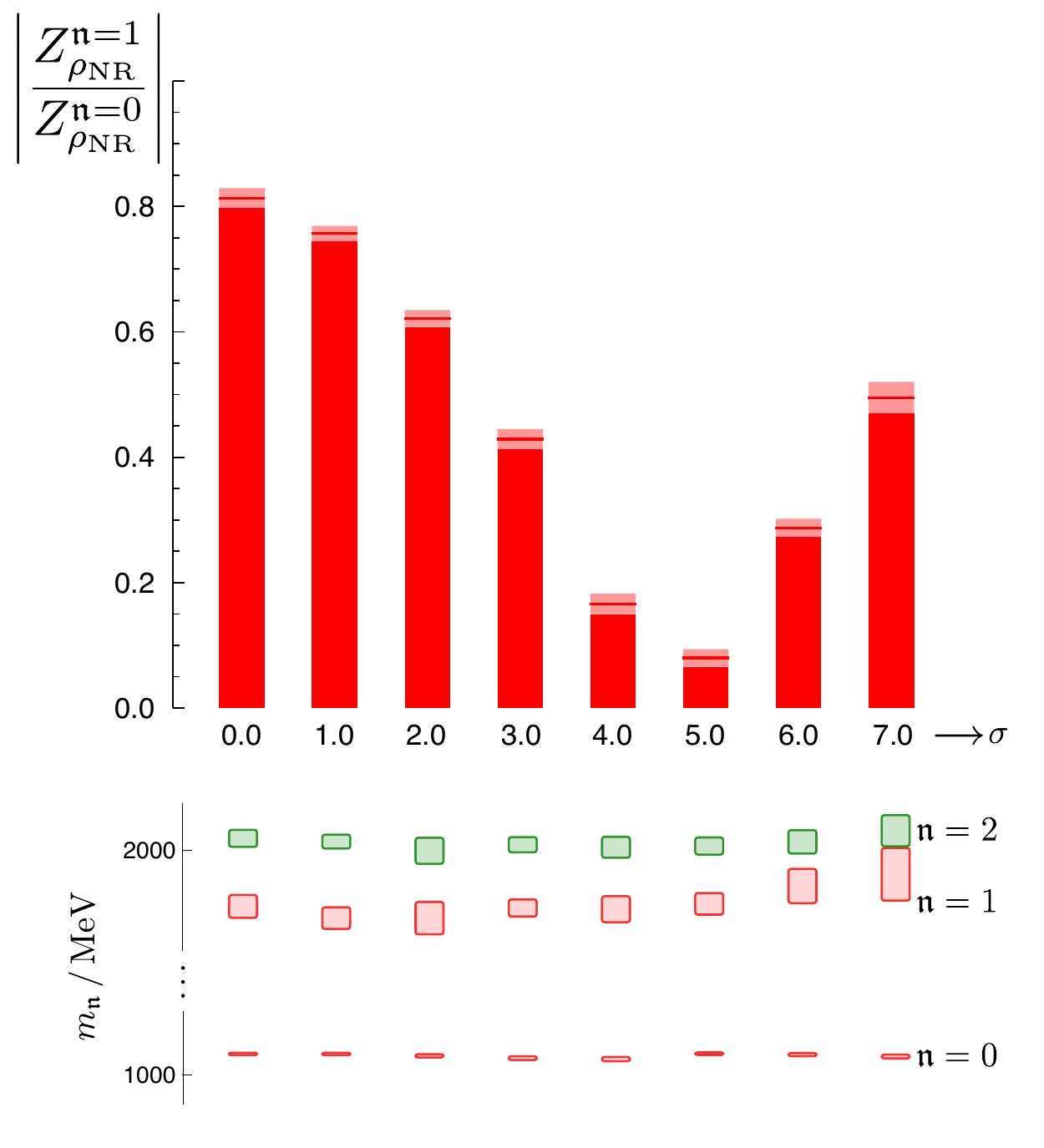}
\caption{Upper panel: Ratio of $1^{--}$ first excited state to ground state overlaps for ``$\rho_\mathrm{NR}$" operator as a function of smearing radius, $\sigma$, in spatial lattice units. Lower panel: Corresponding spectrum of lowest three $1^{--}$ states.\label{smear}}
\end{figure}

Fermion bilinears without derivatives within this model will create constituent $q\bar{q}$ states from the vacuum
\begin{align}
	\int \!\!d^3\vec{x}\; \overline{\widetilde{\psi}^\mathsf{i}} \Gamma \widetilde{\psi}^\mathsf{i}(\vec{x},t=0) \big| 0 \big\rangle &= \int\!\! \frac{d^3 \vec{k}}{(2\pi)^3}\sum_{\sigma=\uparrow\downarrow} \int\!\! \frac{d^3 \vec{\bar{k}}}{(2\pi)^3}\sum_{\bar{\sigma}=\uparrow\downarrow} S_q(|\vec{k}|) S_q(|\vec{\bar{k}}|)\;         \bar{u}_\sigma(\vec{k}) \Gamma v_{\bar{\sigma}}(\vec{\bar{k}})     \int \!\!d^3\vec{x}\; e^{i (\vec{k} + \vec{\bar{k}})\cdot \vec{x} }\;\; \mathsf{b}^{\mathsf{i}\dag}_\sigma(\vec{k}) \mathsf{d}^{\mathsf{i}\dag}_{\bar{\sigma}}(\vec{\bar{k}}) \big| 0\big\rangle \nonumber \\
	&= \int\!\! \frac{d^3 \vec{k}}{(2\pi)^3} \, [S_q(|\vec{k}|)]^2   \sum_{\sigma,\bar{\sigma}=\uparrow\downarrow} \bar{u}_\sigma(\vec{k}) \Gamma v_{\bar{\sigma}}(-\vec{k})\; \Big|  q_\sigma^\mathsf{i}(\vec{k}) \,\bar{q}_{\bar{\sigma}}^\mathsf{i}(-\vec{k})\Big\rangle,
\end{align}
so that rest-frame $q\bar{q}$ meson constructions of the following type will have overlap, 
\begin{align}
\big| q\bar{q}\big(n^{2S+1}L_J\big), m_J \big\rangle = \sum_{m_L, m_S} \big\langle L m_L; S m_S \big| J m_J\big\rangle  \sum_{s,\bar{s}} \big\langle \tfrac{1}{2} s; \tfrac{1}{2} \bar{s}\big|S  m_S \big\rangle  \int\!\! \frac{d^3 \vec{k}}{(2\pi)^3}\, \varphi_{nL}(|\vec{k}|)\; Y_L^{m_L}(\hat{k})\, \Big|  q_s^\mathsf{i}(\vec{k}) \, \bar{q}_{\bar{s}}^\mathsf{i}(-\vec{k}) \Big\rangle.
\end{align}
Here $\varphi$ is the radial momentum-space wavefunction. Which $S,L,J$ combinations have non-zero overlap are determined by which $\Gamma$ is used and can be worked out by expressing the spinors in the Pauli-Dirac basis. 
In all non-zero cases one has the overlap proportional to a radial integral of the form
\begin{equation}
\int k^2 dk\, [S_q(k)]^2\, k^h \, \varphi_{nL}(k),
\end{equation}
and this integral can be accidentally small if $\varphi$ changes sign in the region of integration. We can explore whether this possibility manifests itself in practical calculations by varying the distillation smearing. A possible modified form for the distillation smearing operator acting on quark fields is $\sum_n e^{-\sigma^2 \lambda_n / 4}\, \Theta(\lambda_\mathrm{max} - \lambda_n)\, \xi_n \xi_n^\dag$ where $-\nabla^2 \xi_n = \lambda_n \xi_n$ and where our default choice is $\sigma = 0$. In our bound-state model we will describe this via a smearing function $S_q(k) = e^{-\sigma^2 k^2 / 4} \, \Theta(\Lambda - k)$. 

In a lattice QCD calculation we computed the spectrum of $1^{--}$ isovector states with $m_\pi \sim 700 \,\mathrm{MeV}$ using a large operator basis for a range of smearing radii, $\sigma$, and the lowest 64 eigenvectors of $-\nabla^2$ on a $16^3$ lattice. In Figure \ref{smear} we show the low-lying spectrum extracted, which does not change qualitatively with $\sigma$, along with the ratio of overlaps onto $\rho_\mathrm{NR}$ for the first-excited state and the ground state. We clearly see that for a particular region of smearing radii, $\sigma \sim 4-5$, the overlap of the first excited state is suppressed onto an operator that accurately characterises its angular structure. This is an indication that we must be careful not to rule out a particular state assignment solely on the basis of small overlap values evaluated for one particular smearing function.

\section{Hybrid meson overlaps in a non-relativistic constituent gluon model}

We shall show using an explicit $q\bar{q}g$ state construction that our operators featuring $D^{[2]}_{J=1}$ can interpolate hybrid states featuring a $J_g^{P_gC_g}=1^{+-}$ gluonic state. In order to consider the role of hybrids in the spectrum we must take account of the gauge-field part of the covariant derivatives. As mentioned in the text, the links that enter into discretised derivatives on the lattice are smeared. We will not assume that the corresponding smeared gluonic field is linear in the creation/annihilation of constituent gluons owing to its non-linear construction and the lack of an explicit transformation from Lagrangian gluons to constituent gluons. Our assumed form will be
\begin{align}
\tilde{A}^\mathsf{a}_\mu(x) = \int\!\! \frac{d^3 \vec{q}}{(2\pi)^3} \sum_{\lambda=\pm}\; \frac{S_g(|\vec{q}|)}{\sqrt{2\, \omega(|\vec{q}|)}} \Big[e^{-iqx}\epsilon_\mu(\vec{q},\lambda) \mathsf{a}^\mathsf{a}_\lambda(\vec{q}) + e^{iqx}\epsilon^*_\mu(\vec{q},\lambda) \mathsf{a}^{\mathsf{a}\dag}_\lambda(\vec{q}) + {\cal O}(\mathsf{a}^\dag \mathsf{a}^\dag ) + {\cal O}(\mathsf{a} \mathsf{a} ) +\ldots  \Big] \nonumber.
\end{align}
We will not have cause to go beyond one power of the constituent gluon creation operator in our interpretation of the spectrum.  The constituent gluons created by $\mathsf{a}^{\mathsf{a}\dag}_\lambda(\vec{q})$ have an energy-momentum dispersion $\omega(|\vec{q}|)$ that is not specified, but it is assumed to feature a mass-gap as $|\vec{q}|\to 0$. Since stout-link smearing \cite{Morningstar:2003gk} doesn't change the gauge-transformation properties of the links we'll make the simplifying assumption that the smearing function, $S_g(|\vec{k}|)$ is color-independent and simply acts to damp out high-momentum modes. We have assumed that the constituent gluons have only transverse helicity, $\lambda=\pm 1$, as would be appropriate in Coloumb gauge. The only piece of this decomposition that we will make use of in interpolating hybrid meson states is the term linear in the creation of a constituent gluon.

We shall work under the assumption that a mass gap in the constituent gluon dispersion is such that states with more than one constituent gluon appear high in the spectrum and can be neglected. We construct $q\bar{q}g$ states at rest non-relativistically as in \cite{Guo:2008yz} but couple the internal angular momentum in a slightly different order,
\begin{align}
\Big|q\bar{q}\big(n\,^{2S+1}\!L_{J_{q\bar{q}}}\big) g\big[J_g^{P_g} \big];\; J^{PC}, m_J \Big\rangle	&=
\int \!\!\frac{d^3\vec{q}}{(2\pi)^3}\frac{d^3\vec{p}}{(2\pi)^3}\; \Psi\big(|\vec{q}|, |\vec{p}|\big)\, \sum \nonumber \\ 
&\quad\quad \times \big\langle \tfrac{1}{2}m; \tfrac{1}{2}\overline{m} \big|Sm_S \big\rangle
\big\langle Lm_L; Sm_S \big|J_{q\bar{q}}m_{q\bar{q}} \big\rangle
\big\langle J_{q\bar{q}}m_{q\bar{q}}; J_g m_g \big|Jm_J \big\rangle \nonumber \\
&\quad\quad\quad   \times  Y^{m_L}_L(\hat{p}) \, (-1)^{J_g} \sqrt{\tfrac{2J_g+1}{4\pi}} D^{(J_g)*}_{m_g,-\mu}(-\hat{q})\, \tfrac{1}{\sqrt{2}}\big( \delta_{\mu,+} + \xi \delta_{\mu,-} \big)\nonumber \\
&\quad\quad\quad\quad  \times \tfrac{1}{2}t^\mathsf{a}_{\mathsf{ij}}\, \Big| q^\mathsf{i}_m\big(\!-\!\tfrac{\vec{q}}{2}+ \vec{p}\, \big) \;\bar{q}^{\,\mathsf{j}}_{\overline{m}}\big(\!-\!\tfrac{\vec{q}}{2} - \vec{p}\, \big)\; g^\mathsf{a}_\mu(\vec{q}) \Big\rangle, \label{state}
\end{align}
where the sum is over all $z$-components of angular momenta and helicities, $m ,\overline{m}, m_L ,m_S, m_{q\bar{q}}, m_g, \mu$ and where $\xi = P_g (-1)^{J_g}$. The form of the dependence on the magnitudes of quark and gluon momenta, $\Psi$, will not be specified here. The presence of a generator of $SU(3)$ color, $t^\mathsf{a}_{\mathsf{ij}}$ ensures the $q\bar{q}$ pair are coupled to a color octet. A relativistically invariant construction would need to account for Wigner-rotations arising from the $q\bar{q}$ not being at rest within a $q\bar{q}g$ meson and while these constructions can be performed, \cite{Poplawski:2004qj}, they are somewhat elaborate and we will not attempt them here. 

Operators with a single derivative ($D^{[1]}_{J_D=1}$) are capable of interpolating $q\bar{q}g$ states through the $\tilde{A}$ piece of the derivative, $\overleftrightarrow{D} = \overleftarrow{\partial} - \overrightarrow{\partial} - 2ig \tilde{A}$ as well as $q\bar{q}$ states through the conventional derivative. For $q\bar{q}g$ the overlap will be
\begin{align}
	&\Big\langle q\bar{q}\big(n\,^{2S+1}\!L_{J_{q\bar{q}}}\big) g\big[J_g^{P_g} \big];\; J^{PC}, m_J \Big| \Big(\Gamma \times D^{[1]}_{J_D=1} \Big) \Big| 0 \Big\rangle\nonumber \\
	 &\quad\propto
	\int\!\!\frac{d^3\vec{p}}{(2\pi)^3}\frac{d^3\vec{q}}{(2\pi)^3} \sum \ldots Y_L^{m_L*}(\hat{p})\, \bar{u}_m\big(\!-\!\tfrac{\vec{q}}{2} + \vec{p}\big) \Gamma_{m_\Gamma} v_{\overline{m}}\big(\!-\!\tfrac{\vec{q}}{2} - \vec{p}\big) \ldots D^{(J_g)}_{m_g,-\mu}(-\hat{q}) \epsilon^*_{m_D}(\hat{q},\mu)\, \tfrac{1}{\sqrt{2}}\big( \delta_{\mu,+} + \xi \delta_{\mu,-} \big).
\end{align}
To be somewhat consistent with our neglect of Wigner-rotations in the $q\bar{q}g$ state constructions we will not consider terms in the spinor bilinear $\bar{u}\Gamma v$ that are proportional to the total $q\bar{q}$ momentum, $-\vec{q}$. In this case the integral over $\hat{q}$ can be performed and is non-zero only if $J_g=1$ and $m_g = m_D$. The resulting overlap is proportional to $1+\xi = 1 - P_g$ so that the gluonic excitation interpolated has $J_g^{P_g} = 1^-$. What remains in the overlap are radial integrals and the integral over the $q\bar{q}$ relative momentum direction, $\hat{p}$, which determines which $q\bar{q}$ $L,S$ can contribute.

With two derivatives we can consider four possibilities, 
\begin{align}
D^{[2]} &\sim \big(\, \overleftarrow{\partial}_{\!\!m_1} - \overrightarrow{\partial}_{\!\!m_1} - 2 i g \tilde{A}_{m_1} \big)\big(\, \overleftarrow{\partial}_{\!\!m_2} - \overrightarrow{\partial}_{\!\!m_2} - 2 i g \tilde{A}_{m_2} \big) \quad\quad &    \nonumber \\
&= \overleftarrow{\partial}_{\!\!m_1} \overleftarrow{\partial}_{\!\!m_2} + \overrightarrow{\partial}_{\!\!m_1}  \overrightarrow{\partial}_{\!\!m_2} - \overleftarrow{\partial}_{\!\!m_1} \overrightarrow{\partial}_{\!\!m_2} - \overleftarrow{\partial}_{\!\!m_2} \overrightarrow{\partial}_{\!\!m_1}   &\mathsf{[DD]} \nonumber \\
&\quad - 2ig \big(\, \overleftarrow{\partial}_{\!\!m_1} \tilde{A}_{m_2} - \tilde{A}_{m_2} \overrightarrow{\partial}_{\!\!m_1} + \overleftarrow{\partial}_{\!\!m_2} \tilde{A}_{m_1} - \tilde{A}_{m_1} \overrightarrow{\partial}_{\!\!m_2}  \big) &\mathsf{[DA]} \nonumber \\
&\quad +2ig \Big( \partial_{m_1} \tilde{A}_{m_2} - \partial_{m_2} \tilde{A}_{m_1} + i g \big[\tilde{A}_{m_1}, \tilde{A}_{m_2}\big] \Big) &\mathsf{[F]} \nonumber \\
&\quad- 2 g^2 \big\{ \tilde{A}_{m_1}, \tilde{A}_{m_2} \big\} &\mathsf{[AA]}. \label{DD}
\end{align}
In Table \ref{twoderiv} we present these possibilities projected into definite angular momentum for $q\bar{q}$ and $q\bar{q}g$. Performing the angular integrals in an overlap with states constructed as in Eqn. \ref{state} we find that $D^{[2]}_{J_D=0,2}$ can interpolate $J_g^{P_g}=1^-$ gluonic states while $D^{[2]}_{J_D=1}$ interpolates $J_g^{P_g}=1^+$. The Clebsch-Gordan couplings are such in this model that the overlaps for the lightest hybrid supermultiplet onto the operators $\{\rho_\mathrm{NR}, \pi_\mathrm{NR}\}\times D^{[2]}_{J=1}$ would be expected to have the same value
\begin{align}
\Big\langle q\bar{q}\big(\,^3\!S_1 \big)\, g\big[1^+ \big];\; 0^{-+}\Big|\, \Big(\rho_\mathrm{NR} \times D^{[2]}_{J_D=1} \Big)^{J=0} \,\Big|\, 0 \Big\rangle \nonumber \\
= \Big\langle q\bar{q}\big(\,^3\!S_1 \big)\, g\big[1^+ \big];\; 1^{-+}\Big|\, \Big(\rho_\mathrm{NR} \times D^{[2]}_{J_D=1} \Big)^{J=1} \,\Big|\, 0 \Big\rangle \nonumber \\	
= \Big\langle q\bar{q}\big(\,^3\!S_1 \big)\, g\big[1^+ \big];\; 2^{-+}\Big|\, \Big(\rho_\mathrm{NR} \times D^{[2]}_{J_D=1} \Big)^{J=2} \,\Big|\, 0 \Big\rangle \nonumber \\
= \Big\langle q\bar{q}\big(\,^1\!S_0 \big)\, g\big[1^+ \big];\; 1^{--}\Big|\, \Big(\pi_\mathrm{NR} \times D^{[2]}_{J_D=1} \Big)^{J=1} \,\Big|\, 0 \Big\rangle \nonumber \\
\propto \int \!\! \frac{p^2 dp}{(2\pi)^3} \frac{q^2 dq}{(2\pi)^3}\; \frac{S_g(q)}{\sqrt{\omega(q)}} \,q\, \Psi^*(p,q).
\end{align}
As seen in figure \ref{Z743} this is observed approximately in the lattice data for the states suggested to form the lightest hybrid supermultiplet.

\begin{table}
\begin{tabular}{c|ccc}
$D^{[2]}_{J_D,m_D}$ & $J_D=0$ & $J_D=1$ & $J_D=2$ \\
\hline
$\mathsf{[DD]}(q\bar{q})$ & $|\vec{k}|^2$ & 0 & $|\vec{k}|^2 \, Y_2^{m_D}(\hat{k})$ \\
$\mathsf{[DA]}(q\bar{q}g)$ & $\vec{p} \cdot \vec{\epsilon}^*(\hat{q}, \lambda)$ & 0 & $\langle 1m_1;1m_2|2m_D\rangle p_{m_1} \epsilon^*_{m_2}(\hat{q},\lambda)$ \\
$\mathsf{[F]}(q\bar{q}g)$ & 0 & $\big( \vec{q}\times \vec{\epsilon}^*(\hat{q},\lambda)\big)_{m_D}$ & 0
\end{tabular}
\caption{Angular integrand in overlap of $q\bar{q}$ and $q\bar{q}g$ with $D^{[2]}_{J_D, m_D}$ constructions (as in \ref{DD}) . \label{twoderiv}}
\end{table}

\end{widetext}


\end{document}